\documentclass{aa}  
\usepackage{graphicx}
\usepackage{txfonts}
\usepackage{times}

\newcommand{\Msun}{\mbox{$M_{\odot}$}}

\newcommand{\sub}[1]{\mbox{$_{\rm #1}$}}

\newcommand{\Mhef}{\mbox{$M\sub{Hef}$}}

\newcommand{\Teff}{\mbox{$T\sub{eff}$}}
\newcommand{\logTe}{\mbox{$\log T\sub{eff}$}}
\newcommand{\logL}{\mbox{$\log(L/L_{\odot})$}}
\newcommand{\diff}{\mbox{d}}
\newcommand{\beq}{\begin{equation}}
\newcommand{\eeq}{\end{equation}}
\newcommand{\beqa}{\begin{eqnarray}}
\newcommand{\eeqa}{\end{eqnarray}}
\newcommand{\benu}{\begin{enumerate}}
\newcommand{\eenu}{\end{enumerate}}
\newcommand{\bite}{\begin{itemize}}
\newcommand{\eite}{\end{itemize}}
\newcommand{\bdes}{\begin{description}}
\newcommand{\edes}{\end{description}}

\newcommand{\comment}[1]{}
\begin{document}
\title{{Scaled solar tracks and isochrones in a large region of 
the $Z$--$Y$ plane }
\subtitle{ I. From the ZAMS to the TP-AGB end for $0.15 - 2.5 
    M_{\odot}$ stars }}

\author{G. Bertelli\inst{1} \and L. Girardi\inst{1} \and P. Marigo\inst{2}
\and E. Nasi\inst{1}} 

\offprints{G. Bertelli, \email{gianpaolo.bertelli@oapd.inaf.it}}

\institute {INAF - Padova Astronomical Observatory, Vicolo 
dell'Osservatorio 5, 35122 Padova, Italy 
\and Astronomy Department, Padova University,
 Vicolo dell'Osservatorio 3, 35122 Padova, Italy}

\date{Received / accepted }

% \abstract{}{}{}{}{} 
% 5 {} token are mandatory
 
\abstract
{In many astrophysical contexts, the helium content of stars may differ 
significantly from those usually assumed in evolutionary calculations.}
{In order to improve upon this situation, we have computed tracks and
isochrones in the range of initial masses $0.15 - 20 M_{\odot}$ for a
grid of 39 chemical compositions with the metal content $Z$ between
0.0001 and 0.070 and helium content $Y$ between 0.23 and 0.46. }
{The Padova stellar evolution code has been implemented with updated
physics. New synthetic TP-AGB models allow the extension of stellar
models and isochrones until the end of the thermal pulses along the
AGB. Software tools for the bidimensional interpolation (in $Y$ and
$Z$) of the tracks have been tuned.}
{This first paper presents tracks for low mass stars (from $0.15$ to
$2.5 M_\odot$) with scaled-solar abundances and the corresponding
isochrones from very old ages down to about 1 Gyr. }
{Tracks and isochrones are made available in tabular form for the
adopted grid of chemical compositions in the plane $Z-Y$. As soon as
possible an interactive web interface will allow users to obtain
isochrones of whatever chemical composition and also simulated stellar
populations with different $Y(Z)$ helium-to-metal enrichment laws.  }


\maketitle{}
%
%_____________________________

\section{Introduction}
  
Evolutionary tracks and isochrones are usually computed taking into 
account a fixed law of helium to metal enrichment
$( \Delta Y/\Delta Z )$.   In
fact most of the available tracks follow a linear $Y(Z)$ relation of
the form $Y=Y_P+(\Delta Y/\Delta Z)\,Z$ (e.g. Bertelli et al. 1994
with $Y_{\rm p}=0.23,\Delta Y/\Delta Z=2.5$; or Girardi et al. 2000,
where $Y_{\rm p}=0.23,\Delta Y/\Delta Z=2.25$). Grids of stellar
models in general used $\Delta Y/\Delta Z$ values higher than $\sim 2$ 
in order to fit both the primordial and the solar initial He content.
On one side we recall that WMAP has provided a value of the primordial 
helium ($Y_P\sim 0.248$, Spergel et al. 2003, 2007), significantly higher 
than the value assumed in our previous stellar models. 
On the other side the determinations of the helium enrichment from nearby 
stars and K dwarfs or from Hyades binary systems show a large range of 
values for this ratio (Pagel \& Portinari 1998, Jimenez et al. 2003, 
Casagrande et al. 2007, Lebreton et al. 2001). 
Therefore it is suitable to make available stellar models that can be used 
to simulate different enrichment laws.

Several new results suggest that the naive assumption that the helium 
enrichment law is universal, might not be correct. In fact there has been 
recently evidence of significant variations in the helium content (and 
perhaps of the age) in some globular clusters, that were traditionally 
considered as formed of a simple stellar population of uniform age and 
chemical composition.
 According to Piotto et al. (2005) and Villanova et al. (2007), only 
greatly enhanced helium can explain the color
difference between the two main sequences in $\omega$ Cen. Piotto et
al.(2007) attribute the triple main sequence in NGC 2808 to successive
rounds of star formation, with different helium abundances.  In NGC
2808 a helium enhanced population could explain the MS spread and the HB
morphology ( Lee et al. 2005, D'Antona et al. 2005).

 In general the main sequence morphology can allow to
detect differences in helium content only when the helium content is
really very large. Many globular clusters might have a population with enhanced
helium (up to $Y= 0.3 - 0.33$), but only a few have a small population
with very high helium, such as suggested for NGC 2808.
For NGC 6441 (Caloi \& D'Antona 2007) a very
large helium abundance is required to explain part of the horizontal
branch population, that might be interpreted as due to self-enrichment
and multiple star formation episodes in the early evolution of
globular clusters.
The helium self-enrichment in globular clusters has been discussed in
several papers. Among them we recall Karakas et al. (2006) on the AGB 
self-pollution scenario, Maeder \& Meynet (2006) on rotating low-metallicity
massive stars, while Prantzos and Charbonnel (2006) discuss the drawbacks
of both hypotheses.

 This short summary of recent papers displays
how much interest there is on the enriched helium content of some stellar 
populations. This is the reason why we started these new sets of 
evolutionary models allowing to simulate a large range of helium enrichment
laws.
There are a number of other groups producing extended databases of tracks and 
isochrones both for scaled solar and for $\alpha$-enhanced chemical 
compositions (Pietrinferni et al. 2004, 2006; Yi et al. 2001 and Demarque et
al. 2004; VandenBerg et al. 2006), but their models usually take into account 
a fixed helium enrichment law.  Unlike these authors, very recently Dotter et 
al. (2007) presented new stellar tracks and isochrones for three initial helium
 abundances in the range of mass between $0.1$ and $1.8 \Msun $ for scaled
solar and $\alpha$-enhanced chemical compositions.  

In this paper we present new 
stellar evolutionary tracks for different values of the initial He 
for each metal content. In fact homogeneous  grids of stellar evolution 
models and isochrones in an extended region of the Z-Y plane and for a large 
range of masses  are required 
for the interpretation of photometric and spectroscopic observations of 
resolved and unresolved stellar populations and for the  investigation of the 
formation and evolution of galaxies.

Section 2 describes the input physics and the coverage of the Z-Y plane and 
Section 3  the new synthetic TP-AGB models. Section 4 presents the
evolutionary tracks and relative electronic tables.
Section 5 describes the derived isochrones and the interpolation scheme;
Section 6 presents the comparison with other stellar models databases
and Section 7 the concluding remarks. 

\section {Input physics and coverage of the $Z-Y$ plane}

The stellar evolution code adopted in this work is basically the same
as in Bressan et al. (1993), Fagotto et al. (1994a,b) and Girardi et
al. (2000), but for updates in the opacities, in the rates of energy
loss by plasma neutrinos (see Salasnich et al. 2000), and for the
different way the nuclear network is integrated (Marigo et al. 2001).
In the following we will describe the input physics, pointing out the
differences with respect to Girardi et al. (2000).

\subsection{Initial chemical composition}
\label{sec_chemic}

\begin{table}[!ht]
\caption{Combinations of $Z$ and $Y$ of the computed tracks}
\label{tab_comp}
\smallskip
\begin{center}
{\small
\begin{tabular}{ccccccc}
\hline
\noalign{\smallskip}
Z & Y1 & Y2 & Y3 & Y4 & Y5 & Y6\\
\noalign{\smallskip}
\hline
\noalign{\smallskip}
0.0001 & 0.23 & 0.26 & 0.30 &      & 0.40 &      \\
0.0004 & 0.23 & 0.26 & 0.30 &      & 0.40 &      \\
0.001  & 0.23 & 0.26 & 0.30 &      & 0.40 &      \\
0.002  & 0.23 & 0.26 & 0.30 &      & 0.40 &      \\
0.004  & 0.23 & 0.26 & 0.30 &      & 0.40 &      \\
0.008  & 0.23 & 0.26 & 0.30 & 0.34 & 0.40 &      \\
0.017  & 0.23 & 0.26 & 0.30 & 0.34 & 0.40 &      \\
0.040  &      & 0.26 & 0.30 & 0.34 & 0.40 & 0.46 \\
0.070  &      &      & 0.30 & 0.34 & 0.40 & 0.46 \\
\noalign{\smallskip}
\hline
\end{tabular}
}
\end{center}
\end{table}
 
Stellar models are assumed to be chemically homogeneous when they
settle on the zero age main sequence (ZAMS).  Several grids of stellar
models were computed from the ZAMS with initial mass from $0.15$ to
$20 M_{\odot}$. The first release of the new evolutionary models, presented
in this paper, deals with very low and low mass stars up to
$2.5 M_{\odot}$.  In general for each chemical composition  the separation
in mass is of $0.05 M_{\odot}$ between $0.15$ and $0.6 M_{\odot}$, of $0.1 
M_{\odot}$ between $0.6$ and $2.0 M_{\odot}$.  In addition evolutionary
models at $2.20$ and $2.50 M_{\odot}$ are computed in order to make 
available isochrones  from the Hubble time down to 1 Gyr for each set of 
models.
 The initial chemical composition is in the range $0.0001
\le Z \le 0.070$ for the metal content and for the helium content in
the range $0.23 \le Y \le 0.46$ as shown in Table 1. The
helium content $Y=0.40$ was supplemented at very low metallicities as
useful for simulations of significant helium enrichment in
globular clusters.

For each value of $Z$, the fractions of different metals follow a
scaled solar distribution, as compiled by Grevesse \& Noels (1993) and
adopted in the OPAL opacity tables. The ratio between abundances of
different isotopes is according to Anders \& Grevesse (1989).

\subsection{Opacities}
\label{sec_opac}

The radiative opacities for scaled solar mixtures are from the OPAL
group (Iglesias \& Rogers 1996) for temperatures higher than $\log T =
4$, and the molecular opacities from Alexander \& Ferguson (1994) for
$\log T < 4.0$ as in Salasnich et al. (2000). In the temperature
interval $3.8<\log T < 4.0$, a linear interpolation between the
opacities derived from both sources is adopted. The agreement between
both tables is excellent and the transition is very smooth in this
temperature interval. For very high temperatures ($log T \ge 8.7$)
opacities by Weiss et al. (1990) are used.

The conductive opacities of electron-degenerate matter are from Itoh
et al. (1983), whereas in Girardi et al. (2000) conductive opacities
were from Hubbard \& Lampe (1969). A second difference with respect to
Girardi et al. (2000) lies in the numerical tecnique used to
interpolate within the grids of the opacity tables. In this paper we
used the two-dimensional bi-rational cubic damped-spline algorythm
(see Schlattl \& Weiss 1998, Salasnich et al. 2000 and Weiss \&
Schlattl 2000).

\subsection{Equation of state}
\label{sec_eos}

The equation of state (EOS) for temperatures higher than $10^7$~K is
that of a fully-ionized gas, including electron degeneracy in the way
described by Kippenhahn et al.\ (1967). The effect of Coulomb
interactions between the gas particles at high densities is taken into
account as described in Girardi et al.\ (1996).

For temperatures lower than $10^7$~K, the detailed ``MHD'' EOS of
Mihalas et al.\ (1990, and references therein) is adopted. The
free-energy minimization technique used to derive thermodynamical
quantities and derivatives for any input chemical composition, is
described in detail by Hummer \& Mihalas (1988), D\"appen et al.\
(1988), and Mihalas et al.\ (1988). In our cases, we explicitly
calculated EOS tables for all the $Z$ and $Y$ values of our tracks,
using the Mihalas et al.\ (1990) code, as in Girardi et al. (2000).
We recall that the MHD EOS is critical only for stellar models with
masses lower than 0.7~\Msun, during their main sequence evolution. In
fact, only in dwarfs with masses lower than 0.7~\Msun\ the surface
temperatures are low enough so that the formation of the H$_2$ molecule
dramatically changes the adiabatic temperature gradient. Moreover,
only for masses smaller than $\sim0.4$~\Msun\ the envelopes become
dense and cool enough so that the many non-ideal effects included in
the MHD EOS (internal excitation of molecules, atoms, and ions;
translational motions of partially degenerate electrons, and Coulomb
interactions) become important.

\subsection{Reaction rates and neutrino losses}
\label{sec_rates}

The adopted network of nuclear reactions  involves all the important
reactions of the pp and CNO chains, and the most important
alpha-capture reactions for elements as heavy as Mg (see Bressan et
al.\ 1993 and Girardi et al. 2000). 

The reaction rates are from the compilation of Caughlan \& Fowler
(1988), but for $^{17}{\rm O}({\rm
p},\alpha)^{14}{\rm N}$ and $^{17}{\rm O}({\rm p},\gamma)^{18}{\rm
F}$, for which we use the determinations by 
Landr\'e et al.\ (1990). The
uncertain $^{12}$C($\alpha,\gamma$)$^{16}$O rate was set to 1.7 times
the values given by Caughlan \& Fowler (1988), as indicated by the
study of Weaver \& Woosley (1993) on the nucleosynthesis by massive
stars.The adopted rate is consistent with the recent determination by
Kunz et al. (2002) within the uncertainties.
 The electron screening factors for all reactions are those
from Graboske et al.\ (1973).
The abundances of the various elements are evaluated with the aid of a 
semi-implicit extrapolation scheme, as described in Marigo et al. (2001).

The energy losses by neutrinos are from Haft et al. (1994). Compared with the
previous ones by Munakata et al.\ (1985) used in Girardi et al. (2000), 
neutrino cooling during the RGB is more efficient.

\subsection{Convection}
\label{sec_conv}

The extension of the convective regions is 
determined considering the presence of convective overshoot.
The energy transport in the outer convection zone is described
according to the mixing-length theory (MLT) of B\"ohm-Vitense (1958). The
mixing length parameter $\alpha$ is calibrated by means of the solar model.

{\em Overshoot} The extension of the convective regions takes into account 
overshooting from the borders of both core and envelope convective zones.
In the following we adopt the formulation by Bressan et al. (1981) in which
the boundary of the convective core is set at the layer where the velocity 
(rather than the acceleration) of convective elements vanishes. Also this
non-local treatment of  convection requires a free parameter related to 
 the mean free path $l$ of convective elements through $l=\Lambda_c H_p$,
($H_p$ being the pressure scale height).   The choice of this parameter
determines the extent of the overshooting  
{\em across} the border of the classical core (determined by the Schwarzschild
 criterion).
 Other authors fix the extent
of the overshooting zone at the distance $d=\Lambda_c H_p$ from the border 
of the convective core (Schwarzschild criterion).   

The  $\Lambda_ c$ parameter in Bressan et al.(1981)
formalism is not equivalent to others found in literature. For
instance, the overshooting extension determined by $\Lambda_c =0.5$ with
the Padova formalism roughly corresponds to the   $d=0.25 H_p$ 
 {\em above} the convective border, adopted by the Geneva group
(Meynet et al.\ 1994 and references therein) 
to describe the same physical phenomenum.
This also applies to the extent of overshooting in the case of models 
computed by Teramo group (Pietrinferni et al. 2004), in the Yale-Yonsei 
database (Yi et al. 2001 and Demarque et al. 2004), as they adopted the value
0.2 for the overshoot parameter {\em above} the border of the classical
convective core. In Victoria-Regina stellar models by 
VandenBerg et al. (2006) a free parameter varying with mass is adopted to take 
into account overshoot in the context of Roxburgh's equation for the maximum 
size of a central convective zone, so the comparison is not so straight.
 The non-equivalency of the parameter
used to describe the extension of convective overshooting by different groups
 has been a recurrent source of misunderstanding in the literature.

We adopt the same prescription as in Girardi et al. (2000) for the
parameter $\Lambda_c$ as a function of stellar mass.
$\Lambda_c$ is set to zero for stellar masses $M\le1.0$~\Msun, and 
for $M\ge 1.5 \Msun$, we adopt $\Lambda_c=0.5$,
i.e.\ a moderate amount of overshoting. 	
In the range $1.0<M<1.5 \Msun$, we adopt a gradual increase of the overshoot
efficiency with mass, i.e.\ $\Lambda_c=M/\Msun - 1.0$. This
because the calibration of the overshooting efficiency in this mass 
range is still very uncertain and from observations there are indications 
that this efficiency should be lower than in intermediate-mass stars.

In the stages of core helium burning (CHeB), the value $\Lambda_c=0.5$ is 
used for all stellar masses. This amount of overshooting
significantly reduces the extent of the breathing pulses of convection
found in the late phases of CHeB (see Chiosi et al.\ 1992).

Overshooting at the lower boundary of
convective envelopes is also considered. 
The calibration of the solar model required 
an envelope overshooting not higher than 0.25 pressure scale
height. This value of $\Lambda_e=0.25$ (see Alongi et al.\
1991) was then adopted for the
stars with $0.6\le(M/\Msun)<2.0$, whereas $\Lambda_e=0$ was
adopted for $M \la 0.6\Msun$. 
For $M>2.5\Msun$ a value of $\Lambda_e=0.7$ was assumed as in
Bertelli et al.\ (1994) and Girardi et al. (2000). Finally, for masses 
between 2.0 and $2.5 \Msun$, $\Lambda_e$ was let to increase gradually 
from 0.25 to 0.7, but for the helium burning evolution, where it is always
set to 0.7.

 The adopted approach to overshooting perhaps does not
represent the complex situation found in real stars. However it represents a
pragmatic choice, supported by comparisons with observations. We recall that
 also  the question of
the overshooting efficiency in stars of different masses is still a matter of
debate (see a recent discussion by Claret 2007, and by Deng and Xiong
2007). 
%in astro-ph/0707.0924).

{\em Helium semiconvection} As He-burning proceeds in the convective core of
low mass stars during the early stages of the horizontal branch (HB), 
the size of the convective core increases.
Once the central value of helium falls below $Y_c =0.7$, the 
temperature gradient reaches a local minimum, so that continued overshoot is
no longer able to restore the neutrality condition at the border of the core.
The core splits into an inner convective
core and an outer convective shell. As further helium is captured by the 
convective shell, this latter tends to become stable, leaving behind a region
of varying composition in condition of neutrality.
%($\Delta_R = \Delta_A$)
This zone is called semiconvective.
Shortly, starting from the center and going outwards, the matter
in the radiatively stable region above the formal convective core is mixed 
layer by layer until the neutrality condition is achieved. This picture holds 
during most of the central He-burning phase.

The extension of the semiconvective region
varies with the stellar mass, being important in the low- and intermediate-mass
stars up to about $5 M_\odot$, and negligible in more massive stars.
We followed the scheme by Castellani et al. (1985), as described in Bressan
et al. (1993). 

{\em Breathing convection} As central helium gets as low as $Y_c \simeq 0.1$, 
the enrichment of fresh helium caused by semiconvective mixing enhances the 
rate of energy produced by the $3\alpha$ reactions in such a way that the 
radiative gradient at the convective edge is increased. A very rapid 
enlargement followed by an equally rapid decrease of the fully convective core 
takes place (pulse of convection). Several pulses may occur before the
complete exhaustion of the central helium content. This convective instability
was called breathing convection by Castellani et al. (1985). 
While semiconvection is a true theoretical prediction, the breathing pulses 
are most likely an artifact of the idealized algorithm used to describe mixing
(see Bressan et al. 1993 and references therein). In our code breathing pulses 
are suppressed by imposing that the core cannot be enriched in helium by mixing
with the outer layers more than a fixed fraction F of the amount burnt by
nuclear reactions. In this way a time dependence of convection is implicitely
taken into account.

\subsection{Calibration of the solar model}
\label{sec_sun}

Recent models of the Sun's evolution and interior show that currently observed 
photospheric abundances (relative to hydrogen) must be lower than those of the 
proto-Sun because helium and other heavy elements have settled toward the Sun's
interior since the time of its formation about 4.6 Gyr ago.
The recent update of the solar chemical composition (Asplund et al 2005, 2006 
and Montalban et al. 2006) has led to a decrease in the CNO and Ne abundances 
of the order of $30 \%$. Their new solar chemical composition corresponds to 
the values: $ X=0.7393, Y=0.2485, Z=0.0122$ with $Z/X=0.0165$.
The corresponding decrease in opacity increases 
dramatically the discrepancies between the sound-speed derived from 
helioseismology and the new standard solar model (SSM). Combinations of
increases in opacity and diffusion rates are able to restore part of the 
sound-speed profile agreement, but the required changes are larger than the 
uncertainties accepted for the opacity and the diffusion uncertainties.
Antia and Basu (2006) investigated the possibility of determining the solar 
heavy-element abundances from helioseismic data, used the dimensionless
sound-speed derivative in the solar convection zone and gave a mean value of
$Z=0.0172 \pm 0.002$.

Bahcall et al.(2006) used Monte Carlo simulations to determine the 
uncertainties
in solar model predictions of parameters measured by helioseismology (depth of
the convective zone, the surface helium abundance, the profiles of the sound
speed and density versus radius) and provided determinations of the correlation
coefficients of the predicted solar model neutrino fluxes. They incorporated
both the Asplund et al. recently determined heavy element abundances and the 
higher abundances by Grevesse \& Sauval (1998). Their Table 7 points out that 
the derived characteristic solar model quantities are significantly different
for the two cases.

According to VandenBerg et al. (2007) the new Asplund et al. metallicity for 
the Sun presents some difficulties for fits of solar abundance models to the
M67 CMD, in that they do not predict a gap near the turnoff, which however is 
observed. If the Asplund et al. solar abundances are correct, only those low-Z
solar models that treat diffusion may be able to reproduce the M67 CMD. 

Owing to the many uncertainties,
we adopted for the Sun the initial metallicity of $Z=0.017$, according to
Grevesse \& Sauval (1998). Our choice is a compromise between the previous
solar metal content, Z=0.019 or 0.020, usually considered for evolutionary
models and the significantly lower value supported by Asplund et al. 
(2005, 2006).

The usual procedure for the calibration of the solar model is of computing a 
number of models while varying the MLT parameter  
$\alpha$ and the initial helium content $Y_0$ until the observed solar 
radius and luminosity are reproduced within a predetermined range. Several 
1~\Msun\ models, for different values of the
mixing-length parameter $\alpha$ and helium content $Y_\odot$, are let
to evolve up to the age of 4.6~Gyr. From this series of models, we are
able to single out the pair of $[\alpha, Y_\odot]$ which allows
for a simultaneous match of the present-day solar radius and
luminosity, $R_\odot$ and $L_\odot$.
An additional constraint 
for the solar model comes from the helioseismological
determination of the depth of the outer solar convective zone 
($R_c$ of the order of $0.710-0.716$ $R_\odot$) and from the surface helium 
value.
The envelope overshooting parameter was adopted as $\Lambda_e=0.25$, which
allows a reasonable reproduction of the observed value of $R_c$ (depth of the
solar convective envelope).

Our final solar model reproduces well the solar $R_\odot$, $L_\odot$,
and $R_{\rm c}$ values. The differences with respect to observed values are
smaller than 0.2~\% for $R_\odot$ and $L_\odot$, and $\sim1$~\% for 
$R_{\rm c}$. From
this model with initial $Z_{\odot}=0.017$ we derive the value of initial helium
 $Y_{\odot}=0.260$ and  $\alpha = 1.68$ (this value of the mixing-length
parameter was used in all our stellar models as described previously).

%%%%%%%%%%%%%%%%%%%%%%%%%%%%%%%%%%%%%%%%%%%%%%%%%%%%%%%%%%%%%%%%%%%%%
\subsection {Mass loss on the RGB}
\label{sec_rgbagb}

 Our evolutionary models were computed at constant mass for all stages 
previous to the TP-AGB. However, mass loss by stellar wind during the RGB
of low-mass stars is taken into account at the stage of isochrone 
construction. We use the empirical formulation by Reimers (1975), with the
mass loss rate expressed in function of the free parameter $\eta$, 
by default assumed as 0.35 in our models (see Renzini \& Fusi-Pecci 1988).
The procedure is basically the following: we integrate the mass loss
rate along the RGB of every single track, in order to estimate the actual
mass of the correspondent model of ZAHB. See Bertelli et al. (1994) for
a more detailed description.

This approximation is a good one since the mass loss does not affect 
significantly  the internal structure of models along the luminous part
of the RGB.

%%%%%%%%%%%%%%%%%%%%%%%%%%%%%%%%%%%%%%%%%%%%%%%%%%%%%%%%%%%%%%%%%%%%%

\section{New synthetic TP-AGB models}

An important update of the database of evolutionary tracks is the extension
of stellar models and isochrones until the end of the thermal pulses  along
the Asymptotic Giant Branch, particularly relevant for stellar population
analyses in the near-infrared, where the contribution of AGB stars to the
integrated  photometric properties  is significant.

The new synthetic TP-AGB models
have been computed with the aid of a synthetic code that has been
recently revised and updated in many aspects as described in the paper
by Marigo \& Girardi (2007), to which the reader should refer for
all details.
It should be emphasized that the synthetic TP-AGB model in use does not 
provide a mere analytic description  of this phase, since one key
ingredient is
a complete static envelope model which is numerically integrated from the 
photosphere down to the core. The basic structure of the envelope model
is the same as the one used in the Padova stellar evolution code.  

The major improvements are briefly recalled below.
\begin{itemize}
\item
{\em{Luminosity.}} Thanks to high-accuracy formalisms (Wagenhuber
  \& Groenewegen 1998; Izzard et al. 2004) we follow the complex
  behaviour of the luminosity due to the flash-driven variations
 and the over-luminosity effect caused by hot-bottom burning (HBB).\\
\item
{\em{Effective temperature.}} One fundamental improvement is the
adoption of {\em molecular opacities} coupled to the actual surface
C/O ratio (Marigo 2002), in place of tables valid for scaled solar
chemical compositions (e.g. Alexander \& Ferguson 1994). The most
significant effect is the drop of the effective temperature as soon
as C/O increases above unity as a consequence of the third dredge-up,
in agreement with observations of carbon stars.\\
\item
{\em{ The third dredge-up.}} We use a more realistic description
of the process as a function of stellar mass and metallicity, with the
aid of available analytic relations of the characteristic parameters, i.e.
minimum core mass  $M_{\rm c}^{\rm min}(M,Z)$ and efficiency
$\lambda(M,Z)$, as derived from full AGB calculations (Karakas et al. 2002).\\
\item
{\em{ Pulsation properties.}} A first attempt is made
  to account for the switching between different  pulsation modes
  along the AGB evolution. Basing on available pulsation models for
  long period variables (LPV; Fox \& Wood 1982, Ostlie \& Cox 1986)  we
  derive a criterion on the luminosity to predict the transition from
the first overtone mode to the fundamental one (and viceversa).\\
\item
{\em{ Mass-loss rates.}} These are evaluated with the aid
of different formalisms, based on dynamical atmosphere models for
LPVs, depending not only on stellar mass, luminosity, effective
temperature, and  pulsation period, but also on the chemical type, namely:
Bowen \& Willson (1991) for C/O $<1$, and Wachter et al. (2002) for C/O $>1$.
\end{itemize}

It should be also emphasized that in Marigo \& Girardi (2007)
the free parameters
of the synthetic TP-AGB model -- i.e. the minimum core mass
$M_{\rm c}^{\rm  min}$
and efficiency $\lambda$ --  were calibrated as a function of stellar mass
and metallicity, on the base of two basic observables derived for
both Magellanic Clouds (MC), namely:
i) the counts of AGB stars (for both M- and C-types) in stellar
clusters and ii) the carbon star luminosity functions
in galaxy fields.
As illustrated by Girardi \& Marigo (2007), the
former observable quantifies the TP-AGB lifetimes for both
the oxygen- and carbon-rich phase as a function of stellar mass and
metallicity and it helps to determine $M_{\rm c}^{\rm  min}$. The
latter observable provides complementary constraints to both 
$M_{\rm c}^{\rm  min}$ and $\lambda$ and their
dependence on stellar mass and metallicity.

The calibration carried out by Marigo \& Girardi (2007) 
has been mainly motivated by the aim of 
assuring that for that set of TP-AGB tracks,
the predicted luminosities and lifetimes,
hence the nuclear fuel of this phase, are correctly evaluated.

The same dredge-up parameters as derived by Marigo \& Girardi (2007)
have been adopted for computing the present sets
of TP-AGB tracks, despite of the fact that they 
are chacterized by
different initial conditions at the first thermal pulse
(e.g. $M_{\rm c},\, T_{\rm eff},\, L$),  and  different
chemical compositions of the envelope. This
fact implies that the new sets of TP-AGB tracks are, to a certain extent,
un-calibrated since no attempt was made to have our basic set of
observables reproduced.

However, the choice to keep the same model paramemeters as derived from
the MG07 calibration is still a meaningful option since the present
TP-AGB models a) include already several important improvements in the
treatment of the TP-AGB phase, as recalled at the beginning of this
section; b) they constitute the reference grid from which we will
start the new calibration procedure. The preliminary step will be to
adopt a reasonable enrichment law, i.e. $Y_P$ (zero point) and the
$\Delta Y/\Delta Z$ (slope) so as to limit the calibration of the free
parameters to TP-AGB models belonging to a particular subset of
initial ($Y,\,Z$) combinations. This work is being undertaken.

%qui c'erano tutte le figure
\begin{figure*}
\begin{minipage}{0.45\textwidth} \noindent a)
\resizebox{\hsize}{!}{\includegraphics{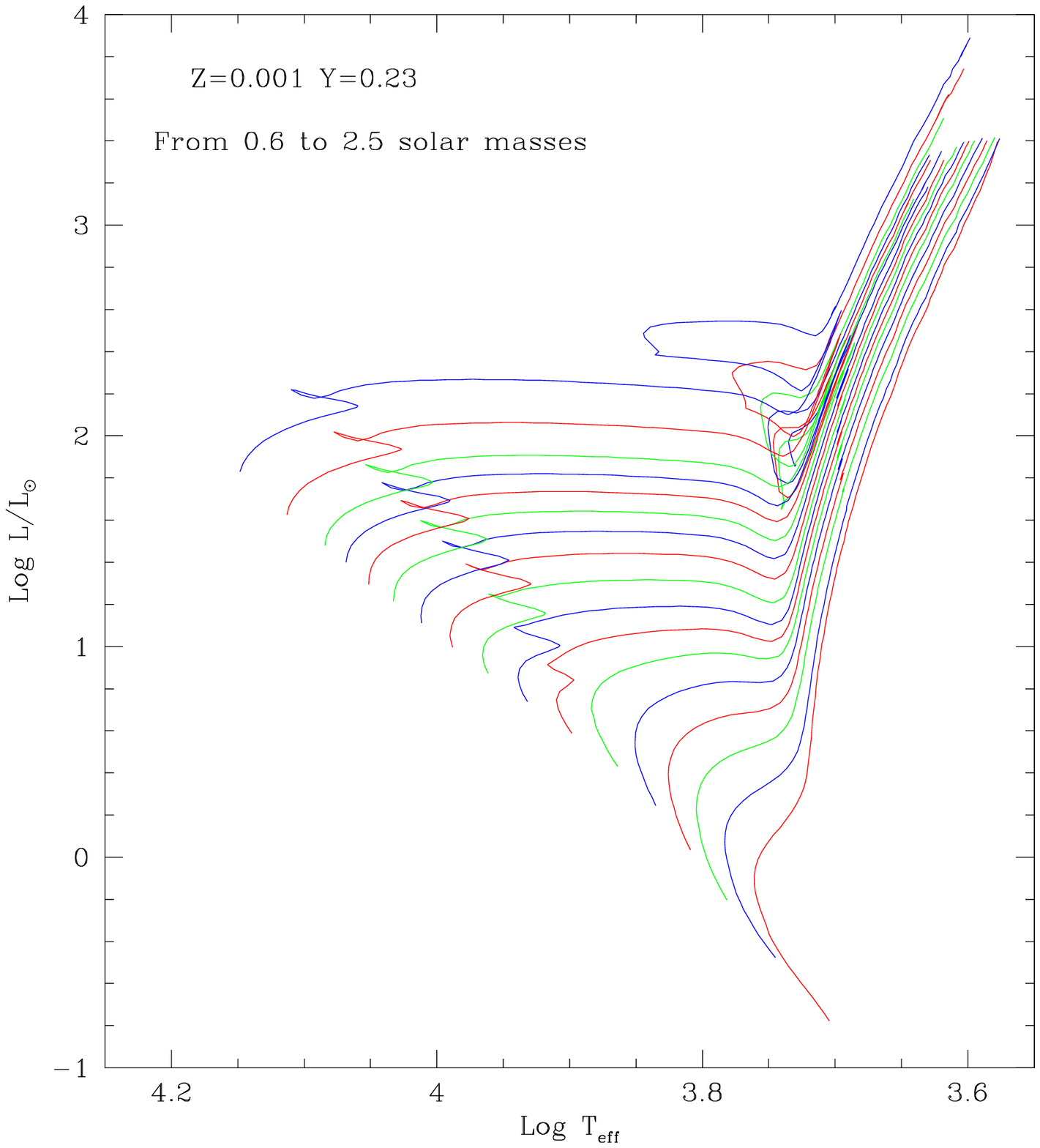}}
\end{minipage} 
\hfill
\begin{minipage}{0.45\textwidth} \noindent b) 
\resizebox{\hsize}{!}{\includegraphics{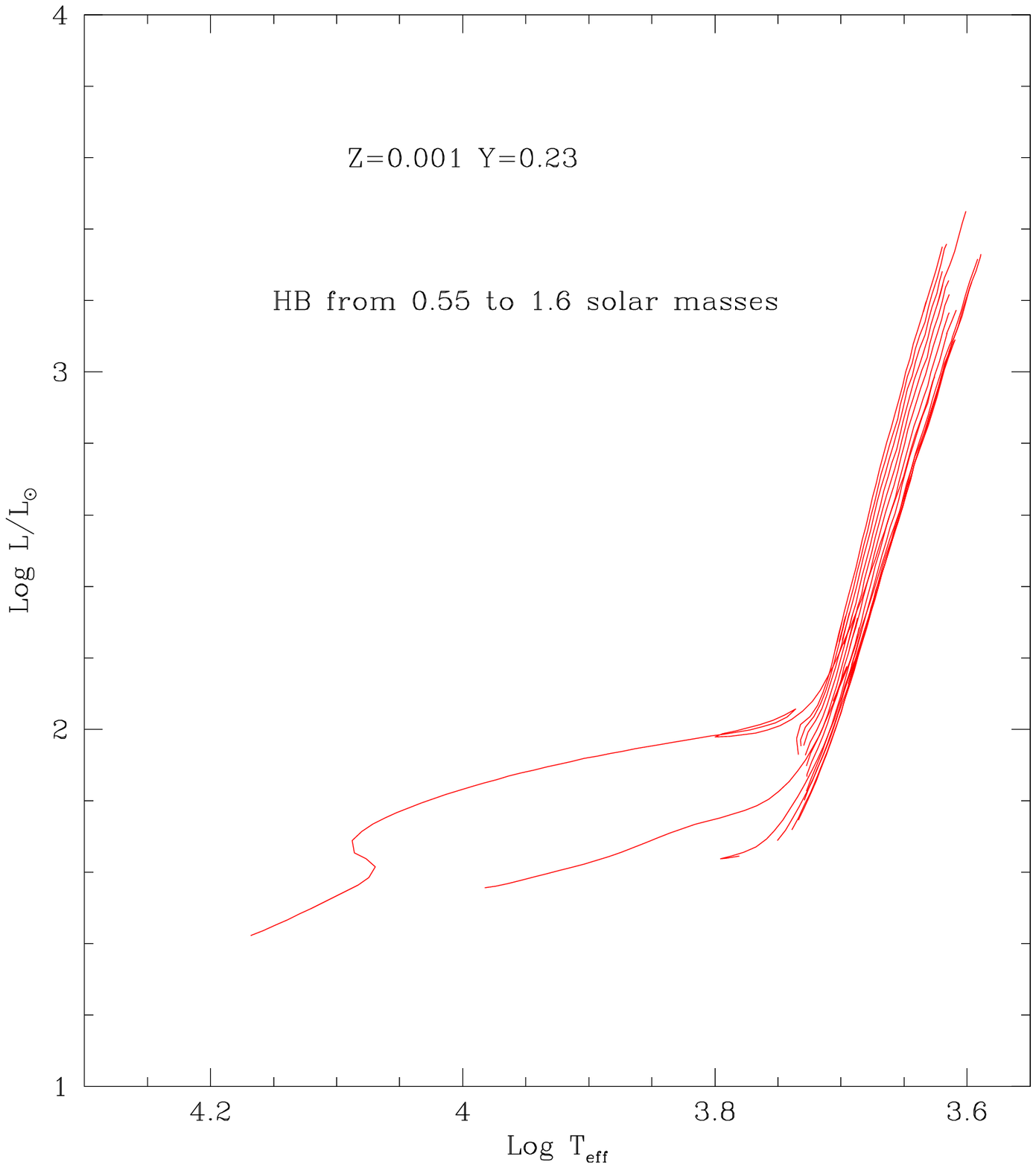}}
\end{minipage} 
\begin{minipage}{0.45\textwidth} \noindent c)
\resizebox{\hsize}{!}{\includegraphics{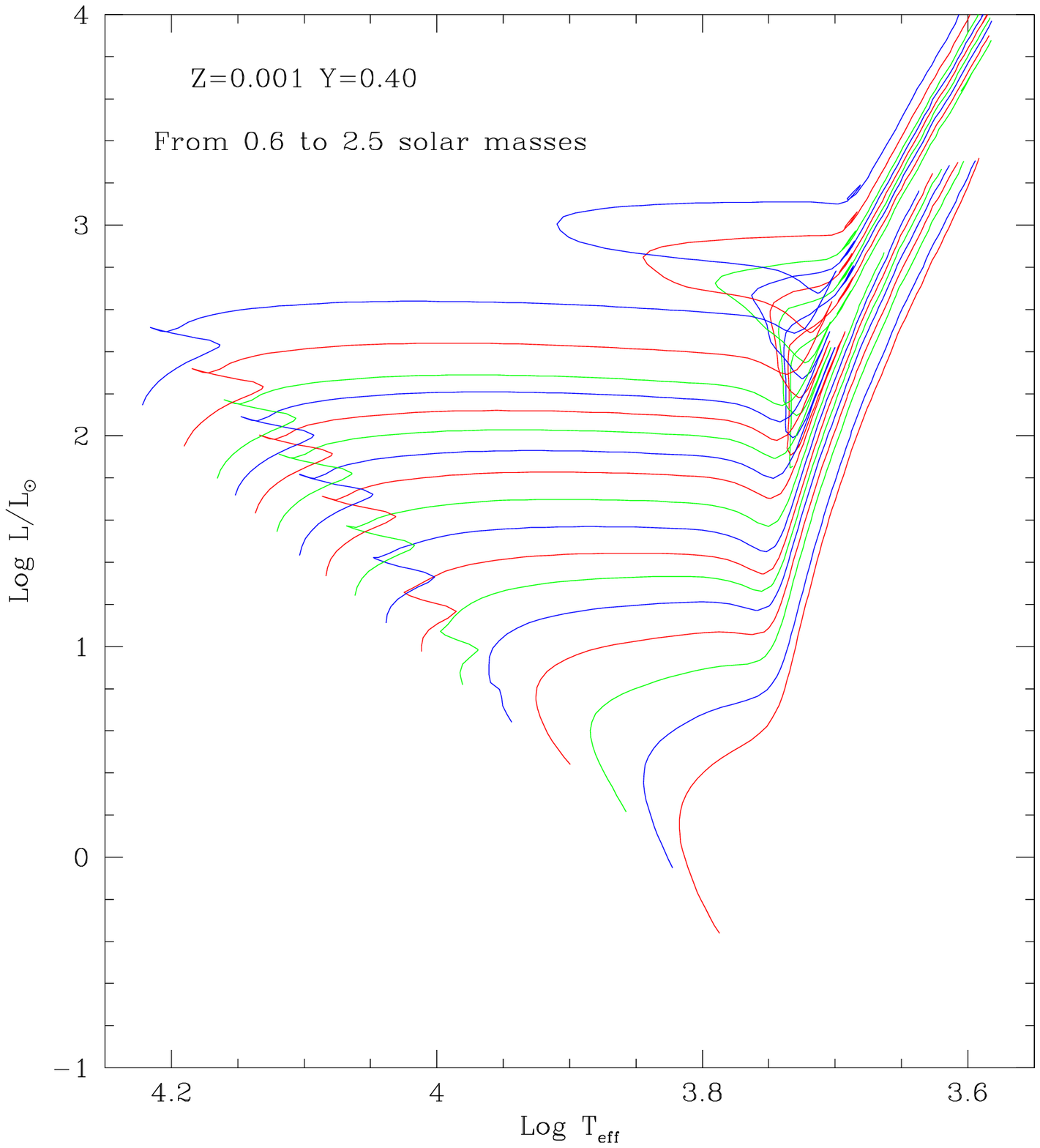}}
\end{minipage} 
\hfill
\begin{minipage}{0.45\textwidth} \noindent d)
\resizebox{\hsize}{!}{\includegraphics{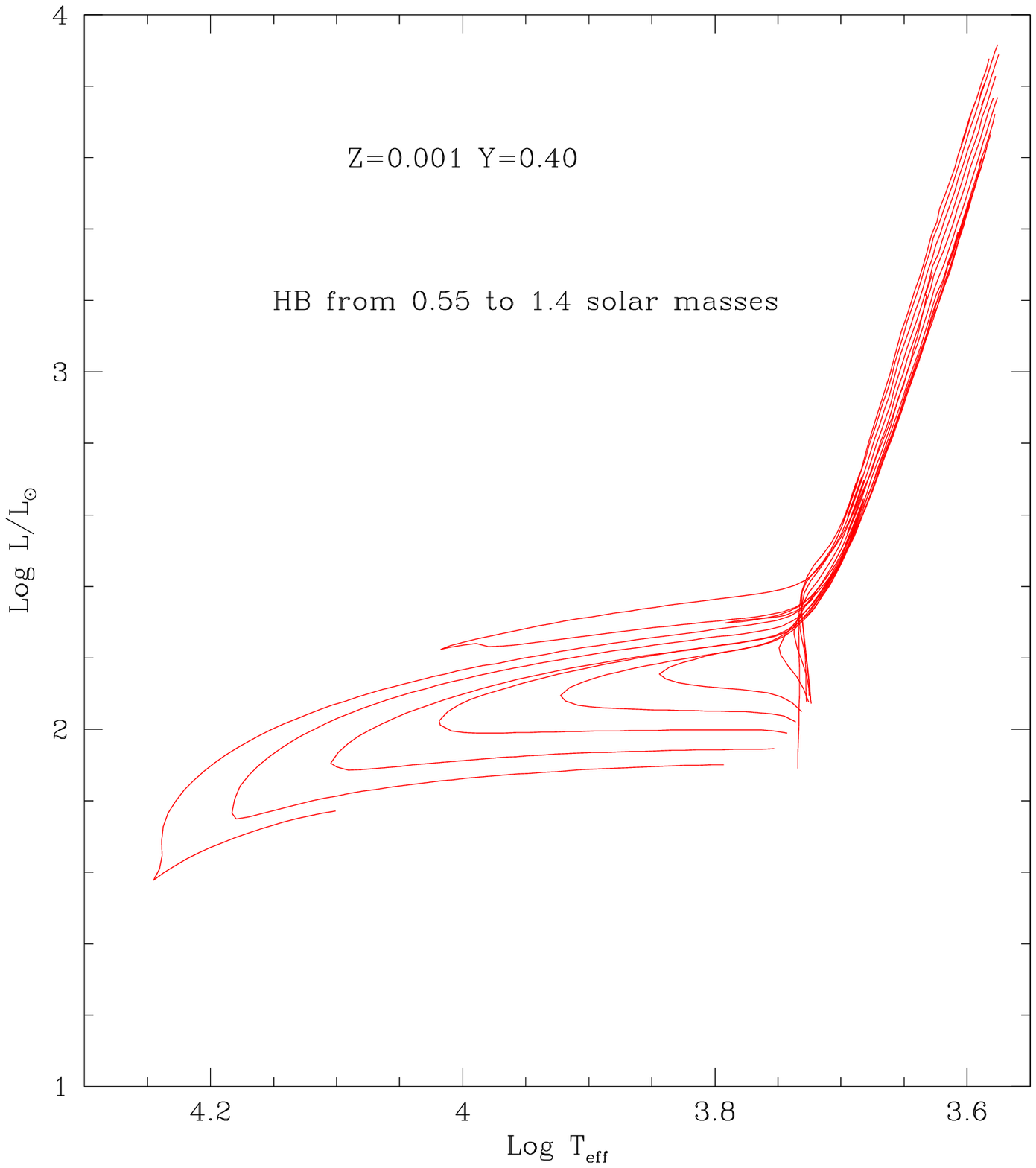}}
\end{minipage} 
\caption{  
Evolutionary tracks in the HR diagram, for the composition $[Z=0.001,
Y=0.23]$. For tracks of low-mass stars up to the RGB-tip 
 and intermediate-mass ones ($M \le 2.5 M_{\odot}$) up to the bTP-AGB 
(panel a), 
 and from the ZAHB up to the bTP-AGB (panel b). In panels c) and d) tracks
are displayed for $[Z=0.001,Y=0.40]$
 }
\label{hrd_z001}
\end{figure*}

\begin{figure*}
\begin{minipage}{0.45\textwidth} \noindent a)
\resizebox{\hsize}{!}{\includegraphics{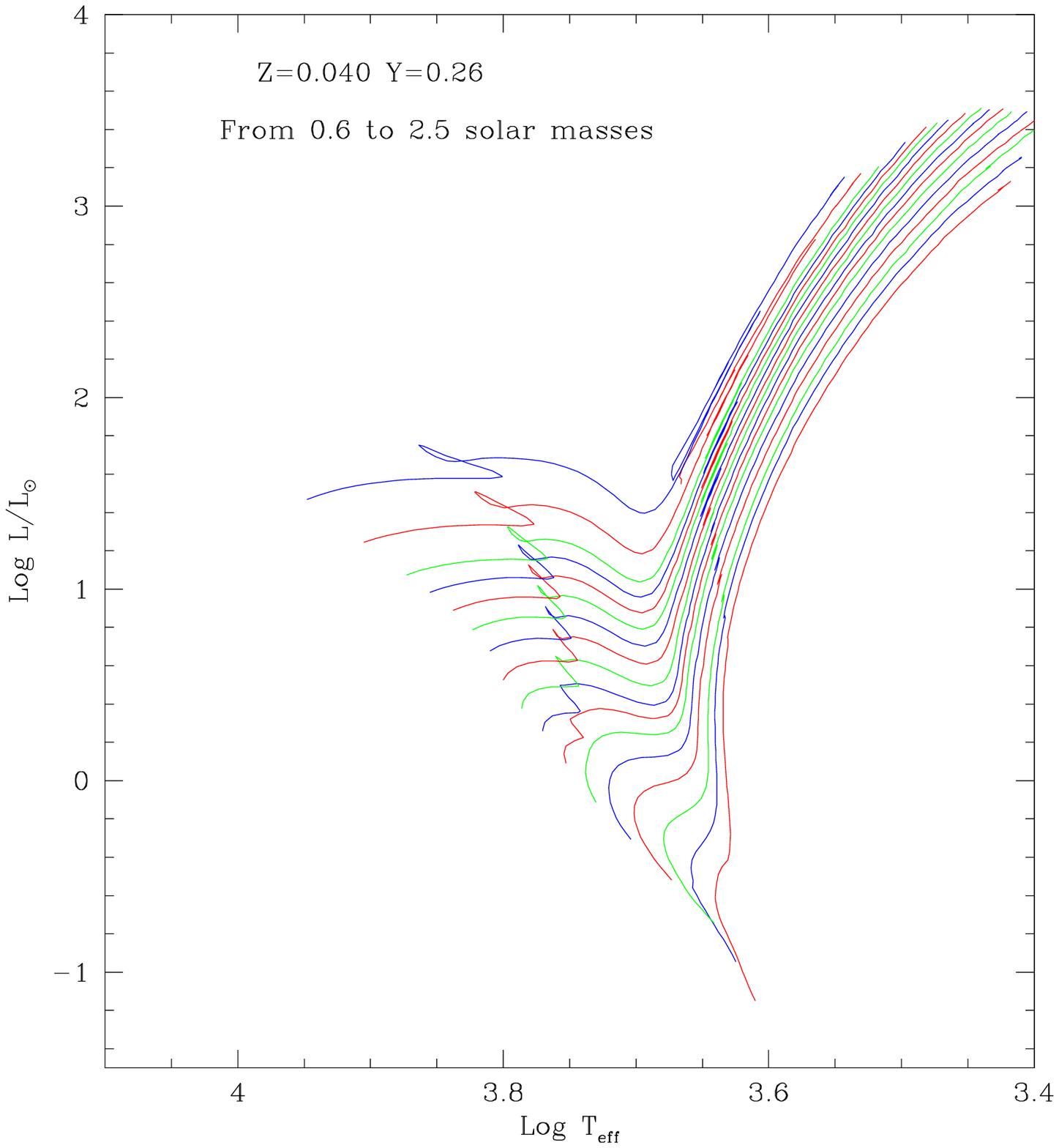}}
\end{minipage} 
\hfill
\begin{minipage}{0.45\textwidth} \noindent b) 
\resizebox{\hsize}{!}{\includegraphics{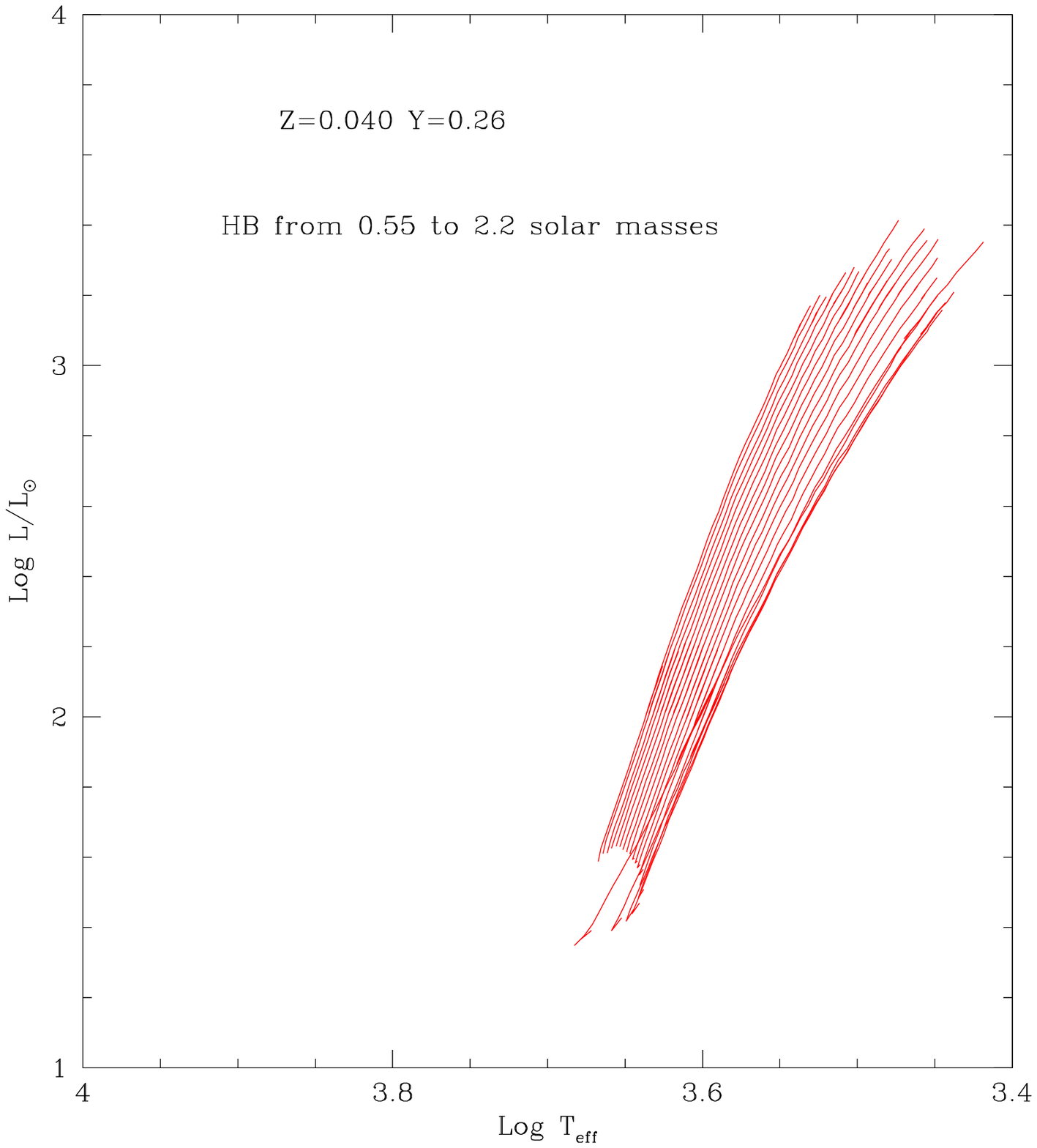}}
\end{minipage} 
\begin{minipage}{0.45\textwidth} \noindent c)
\resizebox{\hsize}{!}{\includegraphics{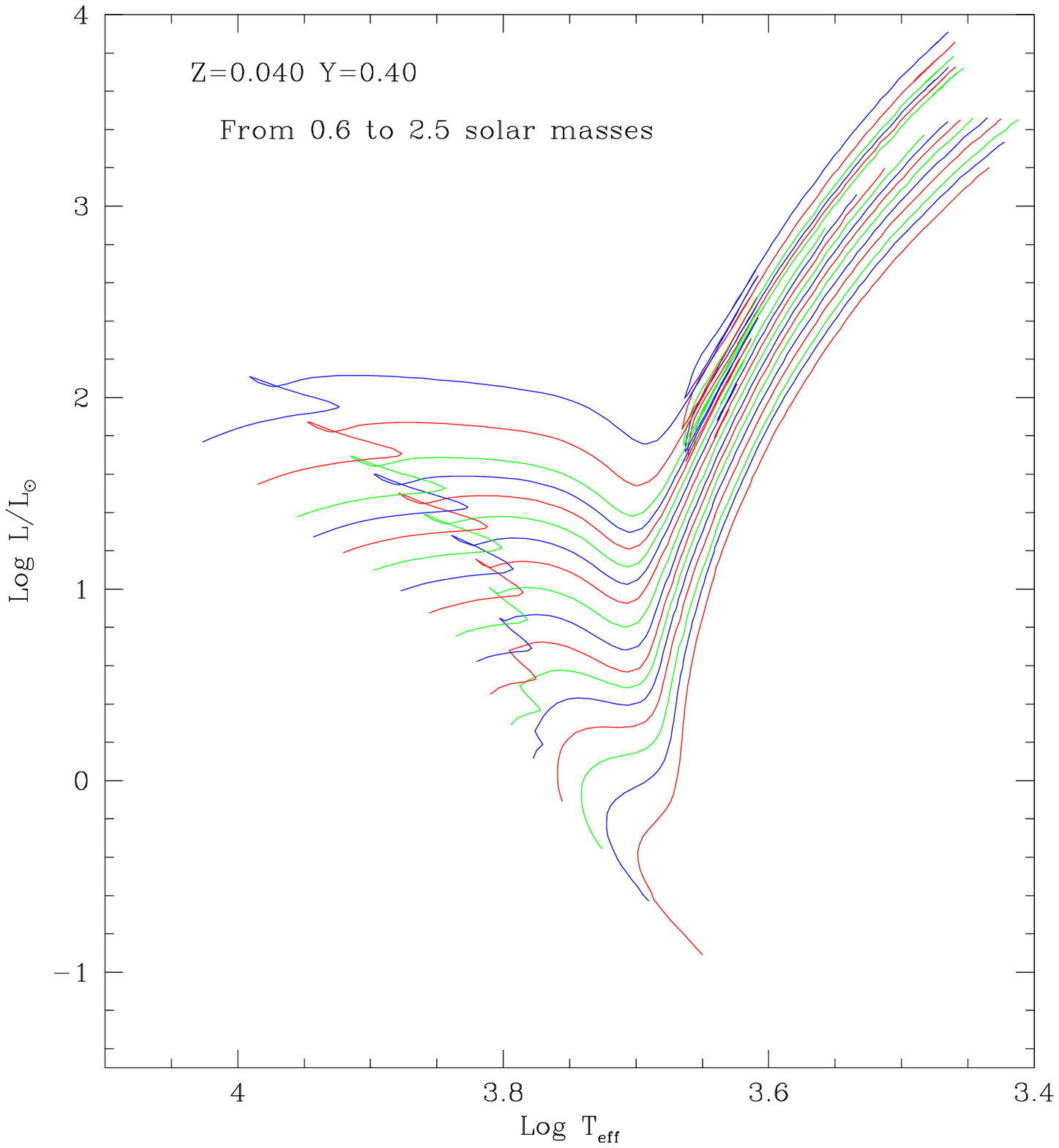}}
\end{minipage} 
\hfill
\begin{minipage}{0.45\textwidth} \noindent d)
\resizebox{\hsize}{!}{\includegraphics{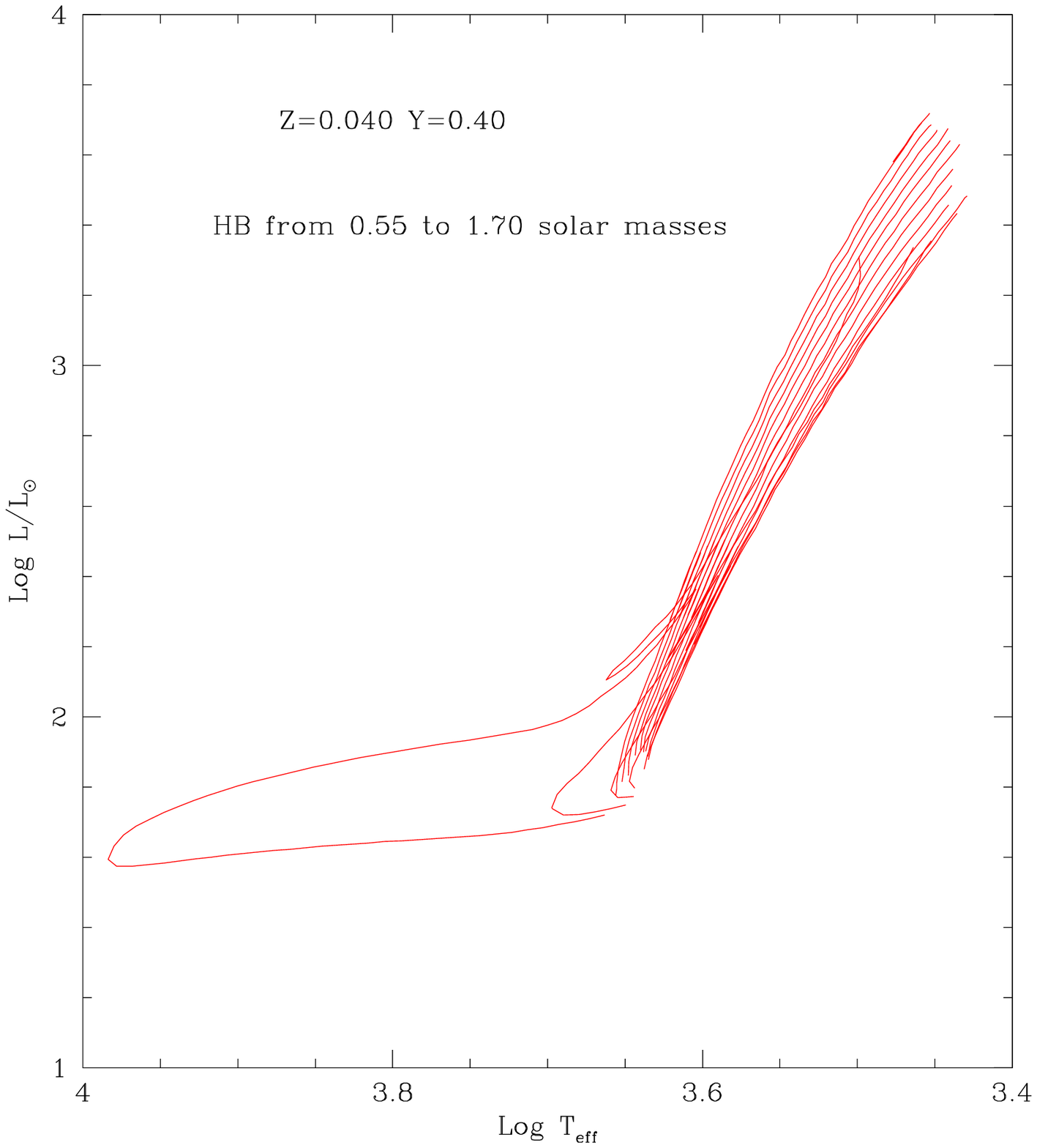}}
\end{minipage} 
\caption{  
Evolutionary tracks in the HR diagram, for the composition $[Z=0.040,
Y=0.26]$. For tracks of low-mass stars up to the RGB-tip 
 and intermediate-mass ones up to the bTP-AGB (panel a), 
 and from the ZAHB up to the bTP-AGB (panels b). In panels c) and d) tracks
are displayed for $[Z=0.040,Y=0.40]$
 }
\label{hrd_z040}
\end{figure*}

\begin{figure*}
\begin{minipage}{0.45\textwidth} \noindent a)
\resizebox{\hsize}{!}{\includegraphics{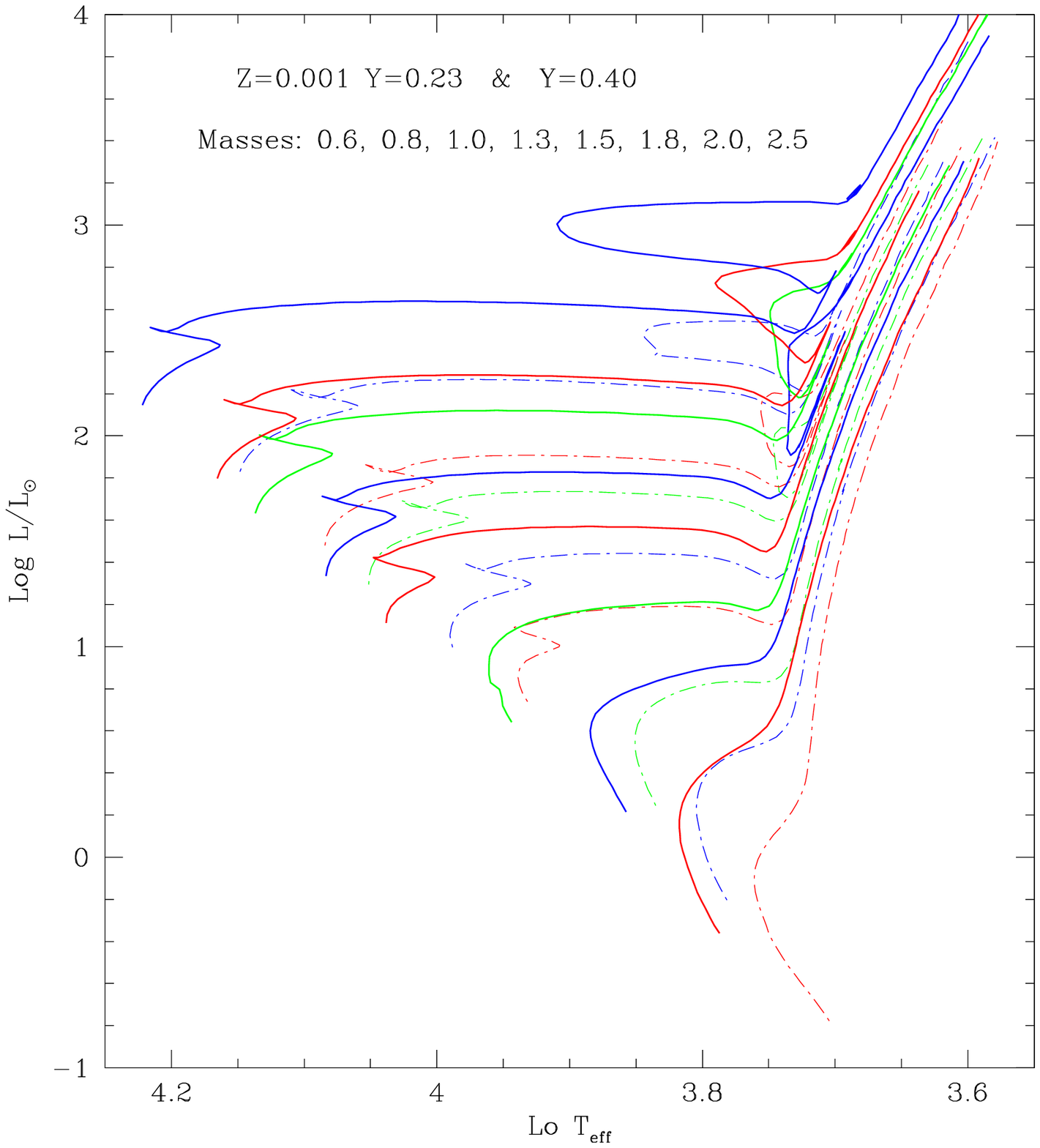}}
\end{minipage} 
\hfill
\begin{minipage}{0.45\textwidth} \noindent b)
\resizebox{\hsize}{!}{\includegraphics{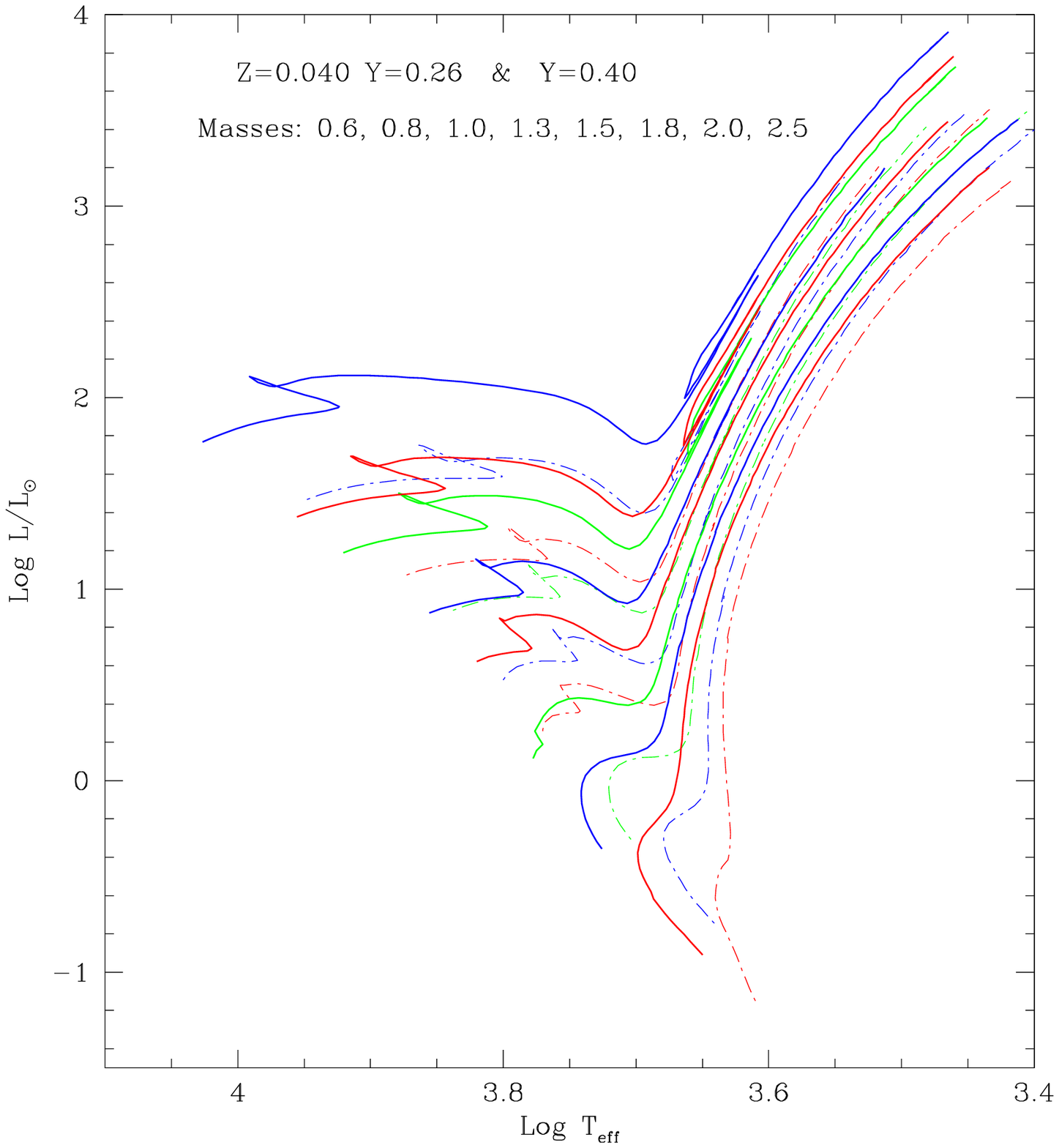}}
\end{minipage} 
\caption{  
Evolutionary tracks in the HR diagram, for the composition $[Z=0.001,
Y=0.23$ (dot-dashed line) and $ Y=0.40$ (solid line)] in panel a). 
 In panel b) tracks are displayed for $[Z=0.040, Y=0.26$ (dot-dashed line) 
 and $Y=0.40$ (solid line)].
 }
\label{hrd_y23y40}
\end{figure*}

%%%%%%%%%%%%%%%%%%%%%%%%%%%%%%%%%%%%%%%%%%%%%%%%%%%%%%%%%%%%%%%%%%%%%
\section{Stellar tracks}
\label{sec_tracks}

\subsection{Evolutionary stages and mass ranges}
\label{sec_massranges}

Our models are evolved from the ZAMS and 
the evolution through the whole
H- and He-burning phases is followed in detail. The
tracks are stopped at the beginning of the TP-AGB phase (bTP-AGB)
in intermediate- and low-mass stars. In the
case of stellar masses lower than 0.6~\Msun, the main sequence
evolution takes place on time scales much longer than a Hubble time. 
For them, we stopped the computations at an age of about 20~Gyr.

In low-mass stars with $0.6 M_{\odot} \le M \le M_{Hef}$ the evolution is 
interrupted at the stage of He-flash in the electron degenerate 
hydrogen-exhausted core.
% This because the computation of the complete 
%evolution through the He-flash requires a huge amount of computing time. 
The evolution is 
then re-started from a ZAHB model with the same core mass and surface
chemical composition as the last RGB model.
For the core we compute the total energy difference between the last RGB
model and the ZAHB configuration and assume it has been provided by helium
burning during the helium flash phase. In this way a certain amount of the 
helium in the core is converted into carbon (about 3 percent in mass, depending
on the stellar mass and initial composition),    
and the initial ZAHB model takes into account the
approximate amount of nuclear fuel necessary to lift the core degeneracy
during the He-flash. The evolution is then followed along the HB up to the 
bTP-AGB phase. We point out that this procedure is more detailed than 
the one adopted in Girardi et al. (2000), where the fraction of He core burnt 
during the He-flash was assumed to be constant and equal to 5\%.

In intermediate-mass stars, the evolution goes from the ZAMS up to
either the beginning of the TP-AGB phase, or to the carbon ignition in 
our more massive models (in this paper we consider masses up to $2.5~\Msun$).

Table 2 gives the values of the transition mass 
\Mhef\, as derived from the present tracks. \Mhef\ is 
the maximum mass for a star to develop an electron-degenerate core 
after the main sequence, and sets the limit between low- and 
intermediate-mass stars (see e.g. Chiosi et al.\ 1992).  
Given the low mass separation between the tracks we computed,
the \Mhef\ values here presented are uncertain by about 0.05~\Msun,

\begin{table}
\caption{The transition mass \Mhef}
\label{tab_mhef}
\begin{tabular}{lllllll}
\noalign{\smallskip}\hline\noalign{\smallskip}
 & \multicolumn{6}{c}{$Y$} \\
 \cline{2-7}
\noalign{\smallskip}
$Z$ & 0.23 & 0.26 & 0.30 & 0.34 & 0.40 & 0.46 \\
\noalign{\smallskip}\hline\noalign{\smallskip}
0.0001 & 1.70 & 1.70 & 1.60 &      & 1.40 &      \\
0.0004 & 1.70 & 1.60 & 1.50 &      & 1.40 &      \\
0.001 & 1.60 & 1.60 & 1.50 &      & 1.40 &     \\
0.002 & 1.70 & 1.60 & 1.50 &      & 1.40 &     \\    
0.004 & 1.80 & 1.75 & 1.60 &      & 1.40 &     \\
0.008 & 1.90 & 1.85 & 1.75 & 1.60 & 1.40 &     \\
0.017 & 2.10 & 2.00 & 1.90 & 1.70 & 1.60 &     \\
0.040 &      & 2.20 & 2.05 & 1.90 & 1.70 & 1.40\\
0.070 &      &      & 2.05 & 1.80 & 1.60 & 1.40\\
\noalign{\smallskip}
\hline
\end{tabular}
\end{table}

\subsection{Tracks in the HR diagram}
\label{sec_hrd}

The complete set of tracks for very low-mass stars ($M<0.6$~\Msun) are 
computed starting at
a stage identified with the ZAMS, and end at the age of 20 Gyr. The ZAMS
model is defined to be the stage of minimum \Teff\ along the computed
track; it follows a stage of much faster evolution in which the pp-cycle 
is out of equilibrium, and in which gravitation may provide a
non negligible fraction of the radiated energy.  
It is well known that these stars evolve very little during the Hubble time.

We display in Figures 1  and 2
only some examples of the sets of 
the computed evolutionary tracks (i.e. $[Z=0.001$ for $Y=0.23$ and $Y=0.40]$ 
and $[Z=0.040$ for $Y=0.26$ and $Y=0.40]$). In these 
figures, panel a) and c)  present the tracks for masses between 0.6 and 
2.5  $M_{\odot}$ from the ZAMS up to the RGB-tip, or to the bTP-AGB, while in
 panel b) and d) the low-mass tracks from the ZAHB up to the bTP-AGB phase are 
plotted from $0.55 M_{\odot}$ to $M_{Hef}$.  
Figure 3 is aimed to illustrate the range of temperatures and 
luminosities involved for the boundary values of helium content at given 
metallicity in the theoretical HR diagram.
The reader can notice, for instance, that at low metallicity and high helium
(Z=0.001, Y=0.040) loops begin to be present during the core He burning phase, 
while they are 
practically missing in the $Z=0.040$ ones for masses about $2.5 M_{\odot}$.

\subsection{Description of the tables}
\label{sec_tabletrack}

The data tables for the present evolutionary tracks are available only
in electronic format.  A website with the complete data-base 
(including additional data and future
extensions) will be mantained at \verb$http://stev.oapd.inaf.it/YZVAR$.

For each evolutionary track, the corresponding data file presents 
23 columns with the following information:
	\benu
	\item \verb$n$: row number;
	\item \verb$age/yr$: stellar age in yr;
	\item \verb$logL$: logarithm of surface luminosity (in solar units), 
\logL;
	\item \verb$logTef$: logarithm of effective temperature (in K), 
\logTe;
	\item \verb$grav$: logarithm of surface gravity ( in cgs units);
	\item \verb$logTc$: logarithm of central temperature (in K);
	\item \verb$logrho$: logarithm of central density (in cgs units);
	\item \verb$Xc$: mass fraction of hydrogen in the stellar centre;
	\item \verb$Yc$: mass fraction of helium in the stellar centre;
	\item \verb$Xc_C$: mass fraction of carbon in the stellar centre;
	\item \verb$Xc_O$: mass fraction of oxygen in the stellar centre;
	\item \verb$Q_conv$: fractionary mass of the convective core;
	\item \verb$Q_disc$: fractionary mass of the first mesh point where 
the chemical composition differs from the surface value;
	\item \verb$L_H/L$: fraction of the total luminosity provided by 
H-burning reactions;
	\item \verb$Q1_H$: fractionary mass of the inner border of the 
H-rich region;
	\item \verb$Q2_H$: fractionary mass of the outer border of the 
H-burning region;
	\item \verb$L_He/L$: fraction of the total luminosity provided by 
He-burning reactions;
	\item \verb$Q1_He$: fractionary mass of the inner border of the 
He-burning region;
	\item \verb$Q2_He$: fractionary mass of the outer border of the 
He-burning region;
	\item \verb$L_C/L$: fraction of the total luminosity provided by 
C-burning reactions;
	\item \verb$L_nu/L$: fraction of the total luminosity lost by 
neutrinos;
	\item \verb$Q_Tmax$: fractionary mass of the point with the highest 
temperature inside the star;
	\item \verb$stage$: label indicating particular evolutionary stages.
\label{item_stage}
	\eenu

A number of evolutionary stages are indicated along the tracks
(column 23). They correspond either to: the
beginning/end of main evolutionary stages, local maxima and minima of
$L$ and \Teff, and main changes of slope in the HR diagram. These
particular stages were, in general, detected by an authomated
algorithm. They can be useful for the derivation of physical
quantities (as e.g.\ lifetimes) as a function of mass and metallicity,
and are actually used as equivalent evolutionary points in our
isochrone-making routines.

For TP-AGB tracks during the quiescent stages of evolution preceding He-shell
flashes, our tables provide:
     \benu
     \item\verb$n$: row number;
     \item\verb$age/yr$: stellar age in yr;
     \item\verb$logL$: logarithm of surface luminosity (in solar units),
\logL;
     \item\verb$logTef$: logarithm of effective temperature (in K),
\logTe;
     \item\verb$Mact$: the current mass (in solar units);
     \item\verb$Mcore$: the mass of the H-exhausted core (in solar units);
     \item\verb$C/O$: the surface C/O ratio.
%\label{item_stage}
     \eenu

\subsection{Changes in surface chemical composition}
\label{sec_chemical}

The surface chemical composition of the stellar models changes 
on two well-defined dredge-up events. The first one occurs at
the first ascent of the RGB for all stellar models (except for 
the very-low mass ones which are not evolved out of the main 
sequence). The second dredge-up is found after the core 
He-exhaustion, being remarkable only in models with 
$M\ga3.5$~\Msun. We provide tables with the surface chemical 
composition of H, $^3$He, $^4$He, and main CNO isotopes, before
and after the first dredge-up, as in this paper we present models
from $0.15$ up to $2.5 M_{\odot}$.
Table 3 shows, as an example, the surface abundances for
the chemical composition Z=0.008 and Y=0.26.

%%%%% Tabella  surf. chem. comp.  z008y26  (Bertelli et al. 2007)

\begin{table*}
\caption{Surface chemical composition (by mass fraction) of 
$[Z=0.008, Y=0.26]$ models.}
\label{tab_du}
\begin{tabular}{lllllllllll}
\noalign{\smallskip}\hline\noalign{\smallskip}
$M/\Msun$ &	 H  &    $^3$He  &      $^4$He  &  $^{12}$C   &    $^{13}$C   &    $^{14}$N   &    $^{15}$N  &     $^{16}$O   &    $^{17}$O  &     $^{18}$O \\
\noalign{\smallskip}\hline\noalign{\smallskip}
\multicolumn{11}{l}{Initial:}   \\
  all   & 0.732 & 2.78$\:10^{-5}$ &  0.260 & 1.37$\:10^{-3}$ & 1.65$\:10^{-5}$ & 4.24$\:10^{-4}$ & 1.67$\:10^{-6}$ & 3.85$\:10^{-3}$ & 1.56$\:10^{-6}$ & 8.68$\:10^{-6}$ \\
\noalign{\smallskip}\hline\noalign{\smallskip}
\multicolumn{11}{l}{After the first dredge-up:} \\

  0.55 &  0.719 & 2.65$\:10^{-3}$ &  0.270 & 1.27$\:10^{-3}$ & 3.99$\:10^{-5}$ & 5.10$\:10^{-4}$ & 1.34$\:10^{-6}$ & 3.85$\:10^{-3}$ & 1.57$\:10^{-6}$ & 8.55$\:10^{-6}$ \\
  0.60 &  0.719 & 2.94$\:10^{-3}$ &  0.270 & 1.26$\:10^{-3}$ & 2.91$\:10^{-5}$ & 5.44$\:10^{-4}$ & 1.41$\:10^{-6}$ & 3.85$\:10^{-3}$ & 1.66$\:10^{-6}$ & 8.31$\:10^{-6}$ \\
  0.65 &  0.715 & 2.34$\:10^{-3}$ &  0.274 & 1.30$\:10^{-3}$ & 3.49$\:10^{-5}$ & 4.91$\:10^{-4}$ & 1.38$\:10^{-6}$ & 3.85$\:10^{-3}$ & 1.59$\:10^{-6}$ & 8.52$\:10^{-6}$ \\
  0.70 &  0.715 & 2.40$\:10^{-3}$ &  0.274 & 1.29$\:10^{-3}$ & 3.27$\:10^{-5}$ & 4.96$\:10^{-4}$ & 1.40$\:10^{-6}$ & 3.85$\:10^{-3}$ & 1.63$\:10^{-6}$ & 8.45$\:10^{-6}$ \\
  0.80 &  0.712 & 1.83$\:10^{-3}$ &  0.278 & 1.26$\:10^{-3}$ & 4.48$\:10^{-5}$ & 5.26$\:10^{-4}$ & 1.31$\:10^{-6}$ & 3.85$\:10^{-3}$ & 1.61$\:10^{-6}$ & 8.46$\:10^{-6}$ \\
  0.90 &  0.709 & 1.43$\:10^{-3}$ &  0.281 & 1.18$\:10^{-3}$ & 4.52$\:10^{-5}$ & 6.13$\:10^{-4}$ & 1.22$\:10^{-6}$ & 3.85$\:10^{-3}$ & 1.68$\:10^{-6}$ & 8.24$\:10^{-6}$ \\
  1.00 &  0.708 & 1.17$\:10^{-3}$ &  0.283 & 1.12$\:10^{-3}$ & 4.55$\:10^{-5}$ & 6.86$\:10^{-4}$ & 1.15$\:10^{-6}$ & 3.85$\:10^{-3}$ & 1.73$\:10^{-6}$ & 7.92$\:10^{-6}$ \\
  1.10 &  0.708 & 9.56$\:10^{-4}$ &  0.283 & 1.07$\:10^{-3}$ & 4.54$\:10^{-5}$ & 7.40$\:10^{-4}$ & 1.09$\:10^{-6}$ & 3.85$\:10^{-3}$ & 1.84$\:10^{-6}$ & 7.64$\:10^{-6}$ \\
  1.20 &  0.710 & 8.20$\:10^{-4}$ &  0.281 & 1.03$\:10^{-3}$ & 4.56$\:10^{-5}$ & 7.90$\:10^{-4}$ & 1.03$\:10^{-6}$ & 3.85$\:10^{-3}$ & 2.25$\:10^{-6}$ & 7.40$\:10^{-6}$ \\
  1.30 &  0.711 & 7.25$\:10^{-4}$ &  0.280 & 9.92$\:10^{-4}$ & 4.60$\:10^{-5}$ & 8.36$\:10^{-4}$ & 9.80$\:10^{-7}$ & 3.85$\:10^{-3}$ & 2.46$\:10^{-6}$ & 7.19$\:10^{-6}$ \\
  1.40 &  0.712 & 6.41$\:10^{-4}$ &  0.279 & 9.50$\:10^{-4}$ & 4.59$\:10^{-5}$ & 8.85$\:10^{-4}$ & 9.25$\:10^{-7}$ & 3.85$\:10^{-3}$ & 3.14$\:10^{-6}$ & 6.94$\:10^{-6}$ \\
  1.50 &  0.713 & 5.71$\:10^{-4}$ &  0.279 & 9.25$\:10^{-4}$ & 4.56$\:10^{-5}$ & 9.19$\:10^{-4}$ & 9.00$\:10^{-7}$ & 3.84$\:10^{-3}$ & 1.12$\:10^{-5}$ & 6.78$\:10^{-6}$ \\
  1.60 &  0.712 & 5.22$\:10^{-4}$ &  0.280 & 9.05$\:10^{-4}$ & 4.54$\:10^{-5}$ & 9.68$\:10^{-4}$ & 8.81$\:10^{-7}$ & 3.80$\:10^{-3}$ & 1.50$\:10^{-5}$ & 6.68$\:10^{-6}$ \\
  1.70 &  0.711 & 4.46$\:10^{-4}$ &  0.281 & 8.92$\:10^{-4}$ & 4.52$\:10^{-5}$ & 1.03$\:10^{-3}$ & 8.63$\:10^{-7}$ & 3.75$\:10^{-3}$ & 1.83$\:10^{-5}$ & 6.57$\:10^{-6}$ \\
  1.80 &  0.709 & 3.99$\:10^{-4}$ &  0.283 & 8.81$\:10^{-4}$ & 4.51$\:10^{-5}$ & 1.09$\:10^{-3}$ & 8.50$\:10^{-7}$ & 3.70$\:10^{-3}$ & 1.59$\:10^{-5}$ & 6.50$\:10^{-6}$ \\
  1.85 &  0.707 & 3.73$\:10^{-4}$ &  0.284 & 8.77$\:10^{-4}$ & 4.44$\:10^{-5}$ & 1.13$\:10^{-3}$ & 8.49$\:10^{-7}$ & 3.66$\:10^{-3}$ & 1.64$\:10^{-5}$ & 6.46$\:10^{-6}$ \\
  1.90 &  0.706 & 3.54$\:10^{-4}$ &  0.285 & 8.77$\:10^{-4}$ & 4.43$\:10^{-5}$ & 1.14$\:10^{-3}$ & 8.46$\:10^{-7}$ & 3.64$\:10^{-3}$ & 1.81$\:10^{-5}$ & 6.44$\:10^{-6}$ \\
  2.00 &  0.706 & 3.19$\:10^{-4}$ &  0.285 & 8.71$\:10^{-4}$ & 4.42$\:10^{-5}$ & 1.17$\:10^{-3}$ & 8.43$\:10^{-7}$ & 3.62$\:10^{-3}$ & 1.47$\:10^{-5}$ & 6.40$\:10^{-6}$ \\
  2.20 &  0.702 & 2.58$\:10^{-4}$ &  0.289 & 8.55$\:10^{-4}$ & 4.44$\:10^{-5}$ & 1.26$\:10^{-3}$ & 8.22$\:10^{-7}$ & 3.54$\:10^{-3}$ & 1.37$\:10^{-5}$ & 6.31$\:10^{-6}$ \\
  2.50 &  0.699 & 2.00$\:10^{-4}$ &  0.292 & 8.42$\:10^{-4}$ & 4.43$\:10^{-5}$ & 1.33$\:10^{-3}$ & 8.07$\:10^{-7}$ & 3.49$\:10^{-3}$ & 1.06$\:10^{-5}$ & 6.21$\:10^{-6}$ \\
 \noalign{\smallskip}\hline\noalign{\smallskip}
\end{tabular}
\end{table*}

\begin{figure}
 \resizebox{\hsize}{!}{\includegraphics{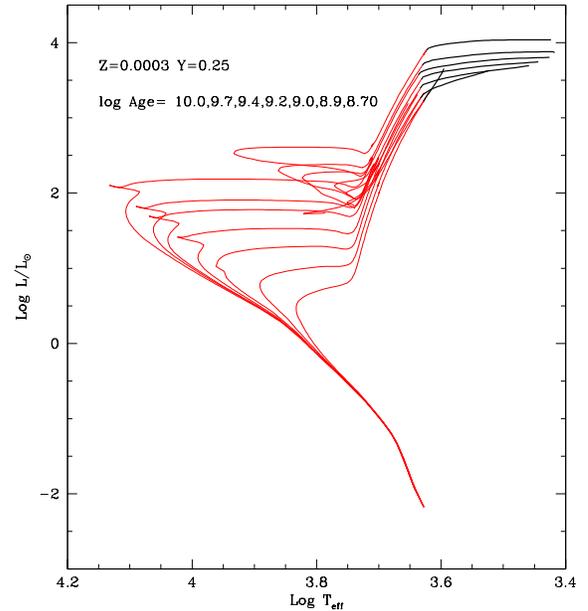}}
\caption{Isochrones for a chemical composition intermediate among those of the
computed tracks (Z=0.0003,Y=0.25) at Log Age= 10, 9.7, 9.4, 9.0, 8.9 and 8.7
years.The new synthetic TP-AGB models allow the extension of the isochrones
(red) until the end of the thermal pulses along the AGB (black). 
}
\label{int_iso1}
\end{figure}

\begin{figure}
 \resizebox{\hsize}{!}{\includegraphics{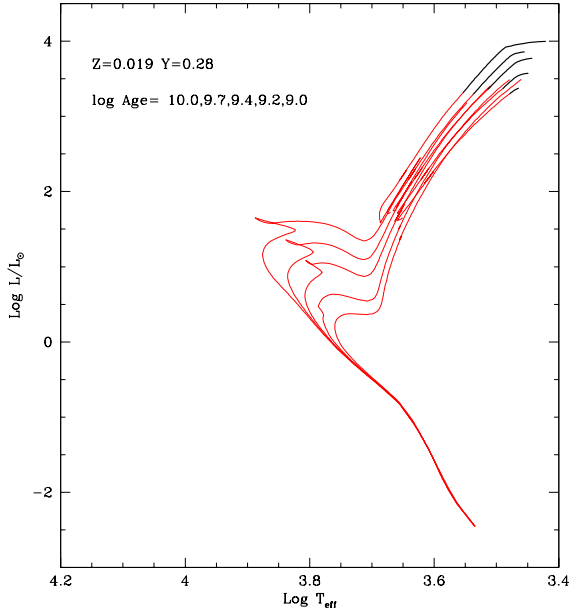}}
\caption{Isochrones for the chemical composition Z=0.019 and Y=0.28,
intermediate among those of the computed tracks at Log Age= 10, 9.7, 9.4, 
9.0 years. The same as Figure 4 for the AGB.
}
\label{int_iso2}
\end{figure}

\begin{figure}
 \resizebox{\hsize}{!}{\includegraphics{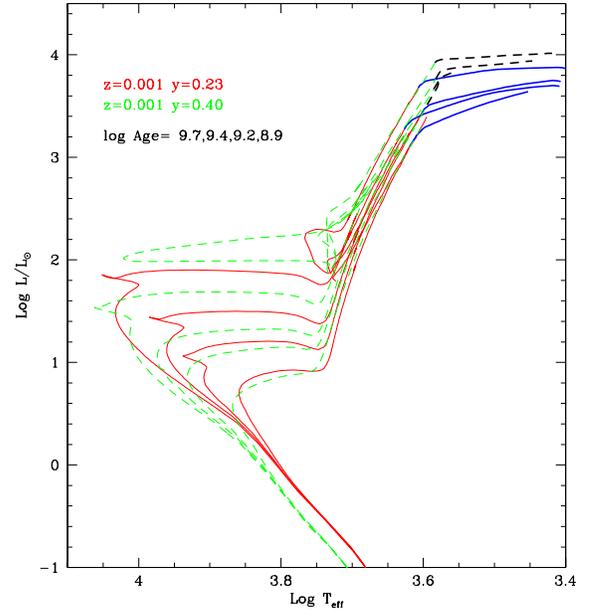}}
\caption{Comparison of isochrones for the same Z=0.001 and different helium 
content. Solid lines correspond to Y=0.23, dashed ones to Y=0.40.
Also in this case the isochrones are extended until the end of the thermal 
pulses along the AGB. }
\label{isozy1y2}
\end{figure}

\section{Isochrones}
\label{sec_isochrones}

From the tracks presented in this paper, we have constructed isochrones
updating and improving the algorithm of ``equivalent evolutionary points'' 
used in Bertelli et al.\ (1994) and Girardi et al. (2000).   
The initial point of each isochrone is the 0.15~\Msun\ model in the 
lower main sequence. The terminal stage of the isochrones is 
the tip of the TP-AGB for the evolutionary tracks presented in this first
paper (up to 2.5 $M_{\odot}$). We recall that usually there are small 
differences in the logarithm of the effective temperature between the last
model of the track after the central He-exhaustion and the starting point
of the synthetic TP-AGB model of the correspondent mass. Constructing the
isochrones we removed these small discontinuities with a suitable shift. 
 The differences arise as the Alexander \& Ferguson (1994) low-temperature
opacity tables, used for the evolutionary models, are replaced by those 
provided by Marigo's (2002) algorithm from the beginning to the end of the 
TP-AGB phase.

In Figures 4 and 5 isochrones are shown for the chemical composition
(Z=0.0003,Y=0.25) and (Z=0.019,Y=0.28)  intermediate
among those of the evolutionary tracks.
The interpolation method to obtain the isochrone for a specific composition
intermediate between the values of the computed tracks is described in the 
following subsection.
In Figures 4, 5 and 6 the plotted isochrones are extended until the end of
the  TP-AGB phase and a different line color points out this phase 
in  the theoretical HR diagram. The flattening of the AGB phase of the 
isochrones marks the transition from M stars to Carbon stars ($C/O > 1$).

An increase of the helium content at the same metallicity in stellar 
models causes a decrease in the mean opacity and an increase in the
mean molecular weight of the envelope (Vemury and Stothers, 1978), and
in turn higher luminosities, hotter effective temperatures and shorter
hydrogen and helium lifetimes of stellar models.
 Apparently in contradiction with the previous statement of 
higher luminosity for helium increase in evolutionary tracks, the evolved
portion of the isochrones with lower helium are more luminous, as shown in 
figure 6 where we plot isochrones with Z=0.001 for Y=0.23 and Y=0.40. 
This effect is related to the 
interplay between the increase in luminosity and the decrease in lifetime 
of stellar models with higher helium content (at the same mass and 
metallicity).

\subsection{Interpolation scheme}

The program ZVAR, already used in many papers (for instance in Ng et
al. 1995, Aparicio et al. 1996, Bertelli \& Nasi 2001, Bertelli et
al. 2003) has been extended to obtain isochrones and to simulate
stellar populations in a large region of the $Z-Y$ plane (now its 
name is YZVAR). 
For each of a few discrete values of the metallicity of former
evolutionary tracks there was only one value of the helium content,
derived from the primordial helium with a fixed enrichment law.  In
the present case we deal with 39 sets of stellar tracks in the
plane $Z-Y$ and we aim at obtaining isochrones for whatever $Z-Y$
combination and stellar populations with the required Y(Z)
enrichment laws.  This problem requires a double interpolation in Y and
Z. We try to describe our method with the help of
Figure 7. In this figure the corners of the box
represent 4 different chemical compositions in a generic mesh of the
grid, i.e. $( Y_1,Z_1)$, $( Y_2,Z_1)$, $( Y_2,Z_2)$, $( Y_1,Z_2)$.

\begin{figure}
 \resizebox{\hsize}{!}{\includegraphics{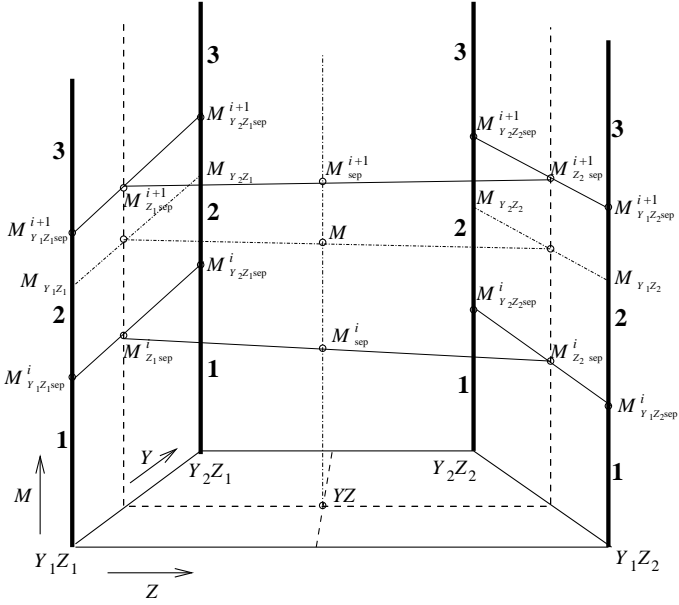}}
\caption{Interpolation scheme inside the range [Z1,Z2] and [Y1,Y2].}
\label{interpolation}
\end{figure}

The vertical lines departing from the corners represent the sequences of
increasing mass of the tracks computed for the corresponding values of 
$(Y,Z)$.  Along each of these lines 
the big points represent the separation masses  between tracks with a
different number of {\em characteristic} points ( a different number of 
equivalent evolutionary phases), indicated as  
 $M_{Y_1Z_1{\rm sep}}^{i}$, $M_{Y_2Z_1{\rm sep}}^{i}$, 
$M_{Y_1Z_2{\rm sep}}^{i}$, $M_{Y_2Z_2{\rm sep}}^{i}$.
The separation masses are marked with the indexes $i$ and $i+1$ in 
Figure 7.
 Actually three intervals of masses are shown in 
this figure, marked with big numbers, identifying the iso-phase intervals,
which means that inside each of them the tracks are 
characterized by the same number of equivalent evolutionary phases
\footnote{With phase we mean a particular stage of the evolution identified 
by common properties both in the C-M diagram and in the internal structure of 
the models.}. The different phases along the tracks are separated by 
{\em characteristic} points, in the following indicated by the index $k$.

For a given mass $M$, age $t$ and  chemical composition $(Y,Z)$ our 
interpolation scheme has 
to determine the luminosity and the effective temperature of the star. This is 
equivalent to assert that we look for the corresponding interpolated track. 
The interpolation must be done inside  the appropriate iso-phase intervals of 
mass. 
For this reason our first step is the identification of the separation masses
($M_{{\rm sep}}^{i},i=1,\ldots,n-1$), where $n$ marks the number of
 iso-phase intervals (n=3 for the case of figure 7). To do this we 
interpolate in $Y$ 
between $M_{Y_1Z_1{\rm sep}}^{i}$ and $M_{Y_2Z_1{\rm sep}}^{i}$ obtaining 
$M_{Z_1{\rm sep}}^{i}$. Analogously we obtain $M_{Z_2{\rm sep}}^{i}$ 
 for the $Z_2$ value. Interpolating  in $\log Z$ between $M_{Z_1{\rm sep}}^{i}$
and $M_{Z_2{\rm sep}}^{i}$  we obtain $M_{{\rm sep}}^{i}$. Analogous procedure 
is followed for all $i=1,\ldots,n-1$ values.

In the example of Fig. 7 the particular mass $M$ that we are 
looking for turns out to be located between $M_{{\rm sep}}^{i}$ and 
$M_{{\rm sep}}^{i+1}$ and in the corresponding  iso-phase interval  labelled
{\bf 2}. Reducing to essentials, we proceed according to the following 
scheme:

\begin{itemize}
\item
Definition of the adimensional mass 

$\tau_m=(M-M_{{\rm sep}}^{i})/(M_{{\rm sep}}^{i+1}-M_{{\rm sep}}^{i})$, 

in this way
$\tau_m$ determines the correspondent masses $M_{Y_1Z_1}$, $M_{Y_2Z_1}$, 
$M_{Y_1Z_2}$, $M_{Y_2Z_2}$  
for each  chemical composition inside the iso-phase interval {\bf 2}.
\item
Interpolation of the tracks $M_{Y_1Z_1}$, $M_{Y_2Z_1}$, $M_{Y_1Z_2}$, 
$M_{Y_2Z_2}$ between pairs of computed tracks (for details see Bertelli et al.
 1990), to obtain
the ages of the characteristic points.
 
%the corresponding luminosities $L_{Y_1Z_1}$, $L_{Y_2Z_1}$, $L_{Y_1Z_2}$, 
%$L_{Y_2Z_2}$.
\item 
Interpolation in $Y$ and $\log Z$ to obtain the ages $t_k$ of the 
characteristic points of the mass M.  We recall that all the masses inside 
the same iso-phase interval show the same number of {\em characteristic}
points. 
\item 
Identification of the current phase  (the actual age $t$ is inside the 
interval ($t_k$,$t_{k+1}$)) and definition of the adimensional time

 $\tau_t= (t-t_k)/(t_{k+1}-t_k)$

\item 
Determination of the luminosities $L_{Y_1Z_1}$, $L_{Y_2Z_1}$,
  etc. relative to the masses $M_{Y_1Z_1}$, $M_{Y_2Z_1}$ etc., 
in correspondence to the adimensional time $\tau_t$ inside the ($k ,k+1$) 
phase.

\item
Interpolation in $Y$ between $L_{Y_1Z_1}$ and $L_{Y_2Z_1}$ obtaining $L_{Z_1}$
, between $L_{Y_1Z_2}$ and $L_{Y_2Z_2}$ obtaining $L_{Z_2}$ and finally 
interpolating in $\log Z$ between $L_{Z_1}$ and $L_{Z_2}$ obtaining the
luminosity $L$ for the mass M and the age $t$.

\item
Analogous procedure to obtain $\log\Teff$ (or other physical properties).
\end{itemize}

For each knot of the grid YZ the data file, used to compute the isochrones,
 contain the evolutionary tracks (log Age, $log L/L_{\odot}$, $log T_{eff}$) 
for the provided range of 
masses and four interfaces (one for each of the four quadrants centered
on the knot). The interfaces are such as to assure, inside every mesh of the
grid, the same number of iso-phase intervals. 
This method is used to obtain an interpolated track of given mass and
chemical composition (inside the provided range) or to derive isochrones of 
given age and chemical composition.

\subsection{Reliability of the interpolation}

\begin{figure}
 \resizebox{\hsize}{!}{\includegraphics{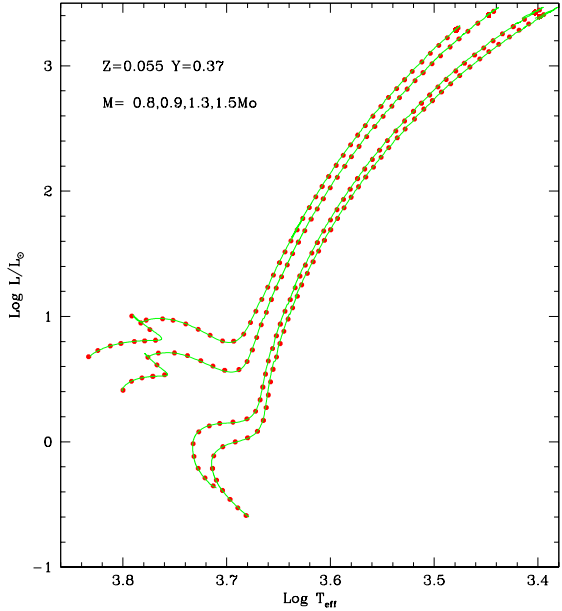}}
\caption{Comparison between computed evolutionary tracks (green line) for a 
chemical composition intermediate among those of the computed grids and the
corresponding interpolated (red points) with the above described method for
 (Z=0.055,Y=0.37)
}
\label{intz055y37}
\end{figure}

\begin{figure}
 \resizebox{\hsize}{!}{\includegraphics{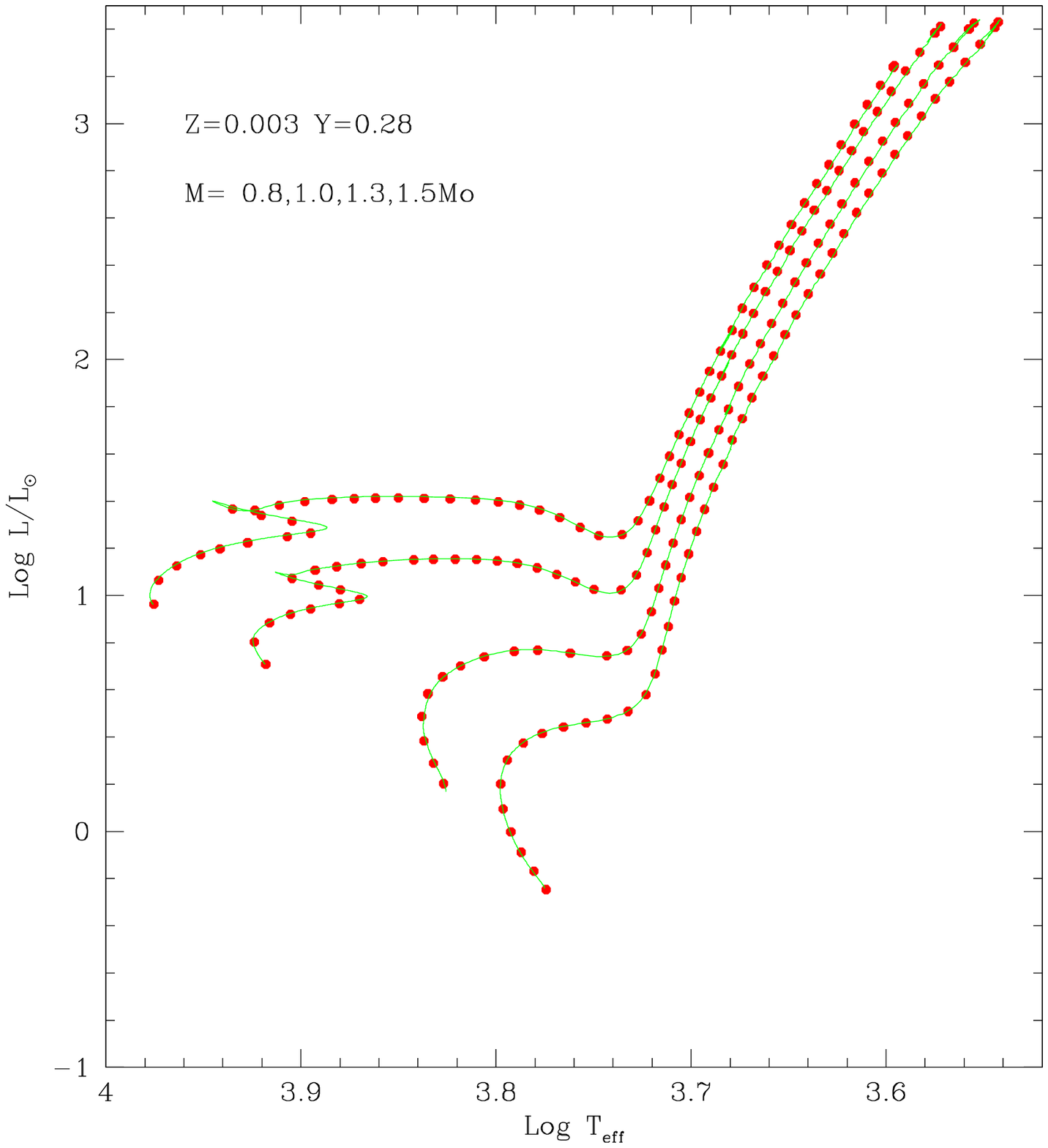}}
\caption{Comparison between computed evolutionary tracks (green line) for a 
chemical composition intermediate among those of the computed grids and the
corresponding interpolated (red points) with the above described method for
 (Z=0.0003,Y=0.28)
}
\label{intz0003y28}
\end{figure}

\begin{figure}
 \resizebox{\hsize}{!}{\includegraphics{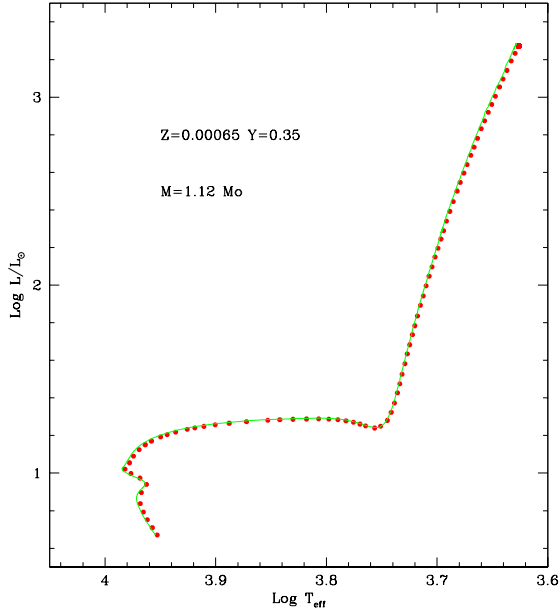}}
\caption{Comparison between a computed evolutionary track for $M = 1.12
M_{\odot}$ (green line) and the correspondent track obtained by interpolation
from the grids (red points), with the chemical composition Z=0.00065 and 
Y=0.35, in the more complex case of close tracks with different morphology
}
\label{intm1_12}
\end{figure}

We have done some tests in order to check the reliability of the
method for the interpolation in the Z-Y plane. For a few combinations
Z-Y we have calculated some tracks and then we have interpolated the
same masses from the existing grids, using the new program YZVAR.  The
results of the comparison are shown in Figs.8, 9 and 10.
In Fig. 8 for the chemical composition Z=0.055 and
Y=0.37 (a point at the center of a mesh) we compare the computed
evolutionary tracks of low mass stars of 0.8, 1.0, 1.3, 1.5 $M_{\odot}$
(green line) and those interpolated with YZVAR (red points). The
result on the HR diagram is satisfactory and the ages differ by less
than $1 \%$.  Fig. 9 presents the results as in Fig 8,
but for the chemical composition Z=0.003 and Y=0.28.  In
Fig. 10 we display a critical case as the considered 1.12
$M_{\odot}$ is between two iso-phase intervals. In fact
this mass is comprised between two masses with different morphology in
the HR diagram. It is evident that the interpolation is able to reproduce
the computed track very well.

\subsection{Bolometric corrections}

The present isochrones are provided in the Johnson-Cousins-Glass
system as defined by Bessell (1990) and Bessell \& Brett (1988). As soon as
possible they will be available in the Vegamag systems of ACS and WFPC2 on 
board of HST (cf. Sirianni et al. 2005; Holtzman et al. 1995) and in the
SDSS system.
The formalism we follow to derive
bolometric corrections in these systems is described in Girardi et
al. (2002). The definition of zeropoints has been revised and is
detailed in a forthcoming paper by Girardi et al. (in prep.; see also
Marigo et al. 2007) and will not be repeated here.

Suffice it to recall that the bolometric correction tables stand
on an updated and extended library of stellar spectral fluxes. The core
of the library now consists of the ``ODFNEW'' ATLAS9 spectral fluxes
from Castelli \& Kurucz (2003), for $T_{\rm eff}$ between 3500 and
50000~K, $\log g$ between $-2$ and $5$, and scaled-solar metallicities
[M/H] between -2.5 and +0.5. This library is extended at the intervals
of high $T_{\rm eff}$ with pure blackbody spectra.
For lower $T_{\rm eff}$, the library is completed
with the spectral fluxes for M, L and T dwarfs from Allard et
al. (2000), M giants from Fluks et al. (1994), and finally the C star
spectra from Loidl et al. (2001).  Details about the implementation of
this library, and in particular about the C star spectra, are provided
in Marigo et al. (2007) and Girardi et al. (in prep.).

It is also worth mentioning that in the isochrones we apply the
bolometric corrections derived from this library without making any
correction for the enhanced He content. As demonstrated in Girardi et
al. (2007), for a given metal content, an enhancement of He 
as high as
$\Delta Y=0.1$ produces changes in the bolometric corrections of just a
few thousandths of magnitude. Just in some very particular situations,
for instance at low $T_{\rm eff}$ and for blue pass-bands, can
He-enhancement produce more sizeable effects on BCs; these situations
however correspond to cases where the emitted stellar flux would
anyway be very small, and therefore are of little interest in
practice.

\subsection{Description of isochrone tables}
\label{sec_tableisoc}
                        
Complete tables with the isochrones can be obtained through the web site 
 \verb$http://stev.oapd.inaf.it/YZVAR$. 
In this data-base, isochrones are provided at $\Delta\log t=0.05$
intervals; this means that any two consecutive isochrones differ by
only 12 percent in their ages.

For each isochrone table the corresponding data file presents 
16 columns with the following information:
	\begin{description}
	\item \verb$1. logAge$: logarithm of the age in years;
	\item \verb$2. log(L/Lo)$: logarithm of surface luminosity (in solar units);
	\item \verb$3. logTef$: logarithm of effective temperature (in K);
	\item \verb$4. logG$: logarithm of surface gravity (in cgs units);
        \item \verb$5. Mi$: initial mass in solar masses;
        \item \verb$6. Mcur$: actual stellar mass in solar masses;  
        \item \verb$7. FLUM$: indefinite integral over the initial mass M of 
the Salpeter initial mass function by number;
        \item \verb$8. - 15.$: UBVRIJHK absolute magnitudes in the 
Johnson-Cousins-Glass system; 
        \item \verb$16. C.P.$: index marking the presence of a characteristic 
point, when different from zero.
	\end{description}

 We recall that the
initial mass is the useful quantity for population synthesis calculations,
since together with the initial mass function it determines the relative 
number of stars in different sections of the isochrones.
 In  column 7  the 
indefinite integral over the initial mass $M$ of the initial mass 
function (IMF) by number, 

i.e.\ 
	\begin{equation}
\mbox{\sc flum} = \int\phi(M) \diff M
	\end{equation}
is presented, for the case of the Salpeter IMF, $\phi(M)=AM^{-\alpha}$, 
with $\alpha=2.35$. When we assume a normalization constant of $A=1$, 
{\sc flum} is simply given by {\sc flum}$ = M^{1-\alpha}/(1-\alpha)$.
This is a useful quantity since the difference between any two 
values of {\sc flum} is proportional to the number of stars located in 
the corresponding mass interval. It is worth remarking that we 
present {\sc flum} values 
for the complete mass interval down to 0.15~\Msun, always assuming a
Salpeter (1955) IMF, whereas we know that such an IMF cannot be extended to
such low values of the mass. However, the reader can easily derive 
{\sc flum} relations for alternative choices of the IMF, by using 
the values of the initial mass we present in Column 5 of the
isochrone tables.
In the last column when there appears the value 1, it marks the presence 
of a characteristic evolutionary point from the ZAMS to the beginning of
the early AGB phase, while the value 2 is related to the characteristic 
points of the TP-AGB phase. 
If there is only one characteristic point (marked with 2) at the 
end of the isochrone, this means that the TP-AGB phase is very short.
The beginning and the end of the TP-AGB phase are pointed out with index 2,
as well as when C/O increases above unity (transition from M to carbon stars).
  
\section{Comparison with other databases}

\begin{figure}
 \resizebox{\hsize}{!}{\includegraphics{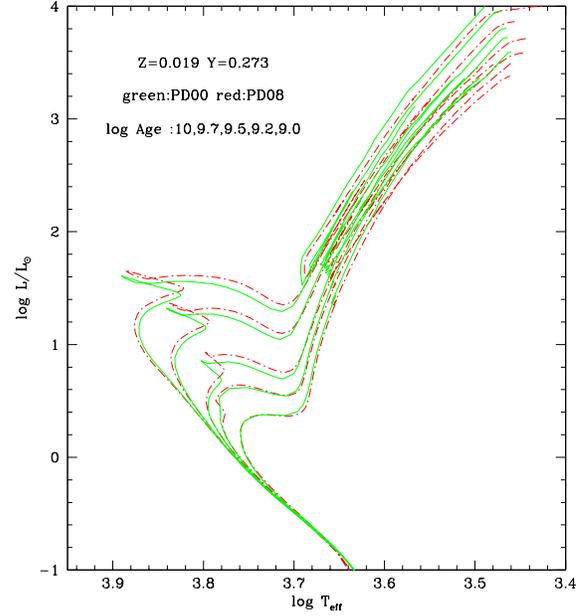}}
\caption{Comparison between the new Padova isochrones (PD08) and Girardi et al.
(PD00) ones for Z=0.019 and Y=0.273 and log Age=10., 9.7, 9.5, 9.2, 9. years.
Solid lines correspond to PD00, dash-dotted ones to PD08 isochrones.}
\label{compPD00_07}
\end{figure}

\begin{figure*}
\begin{minipage}{0.45\textwidth} \noindent a)
\resizebox{\hsize}{!}{\includegraphics{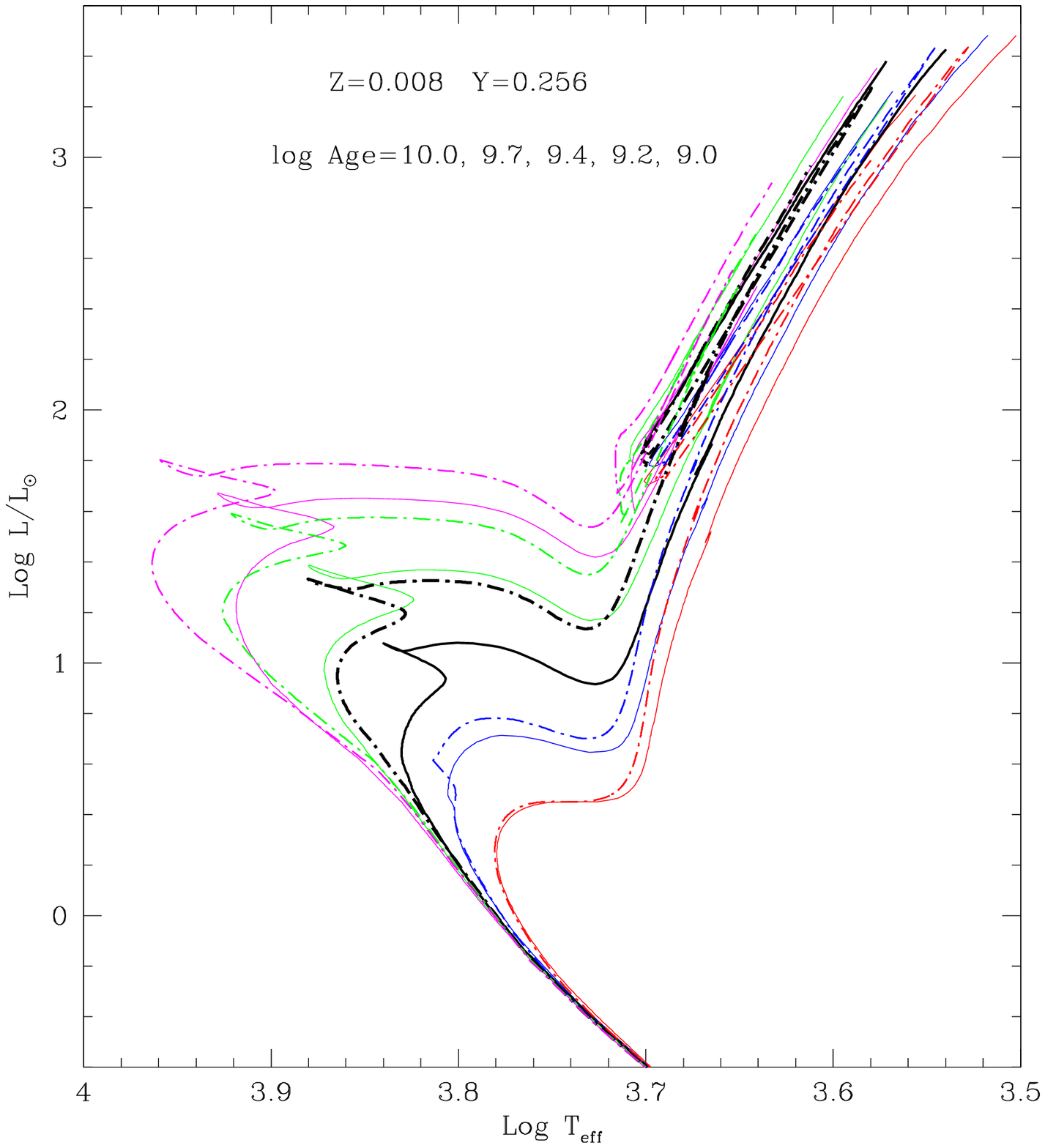}}
\end{minipage} 
\hfill
\begin{minipage}{0.45\textwidth} \noindent b)
\resizebox{\hsize}{!}{\includegraphics{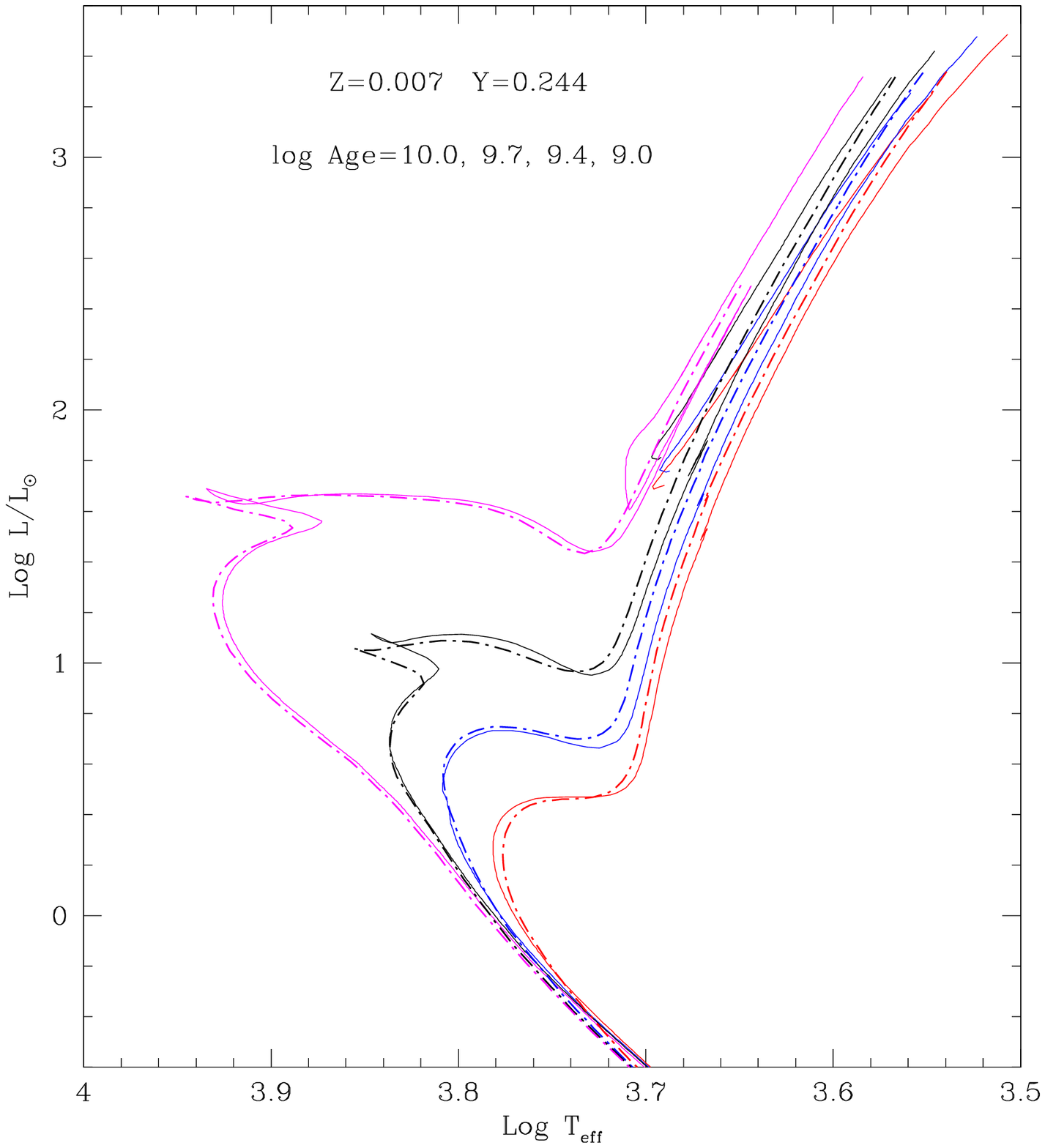}}
\end{minipage} 
\caption{  
Comparison between Padova and Teramo isochrones with overshoot for the 
composition $[Z=0.008, Y=0.256]$ in panel a). Dot-dashed lines correspond to
Teramo ones  and solid lines to our new isochrones. The largest difference
between the PD and Te isochrones is met at log Age=9.4.  
In panel b) the comparison is between Padova (solid line) and $YY$ isochrones 
(dot-dashed) with  Z=0.007 and Y=0.244  }
\label{isoPD_TE_Y2}
\end{figure*} 

\begin{figure*}
\begin{minipage}{0.45\textwidth} \noindent a)
\resizebox{\hsize}{!}{\includegraphics{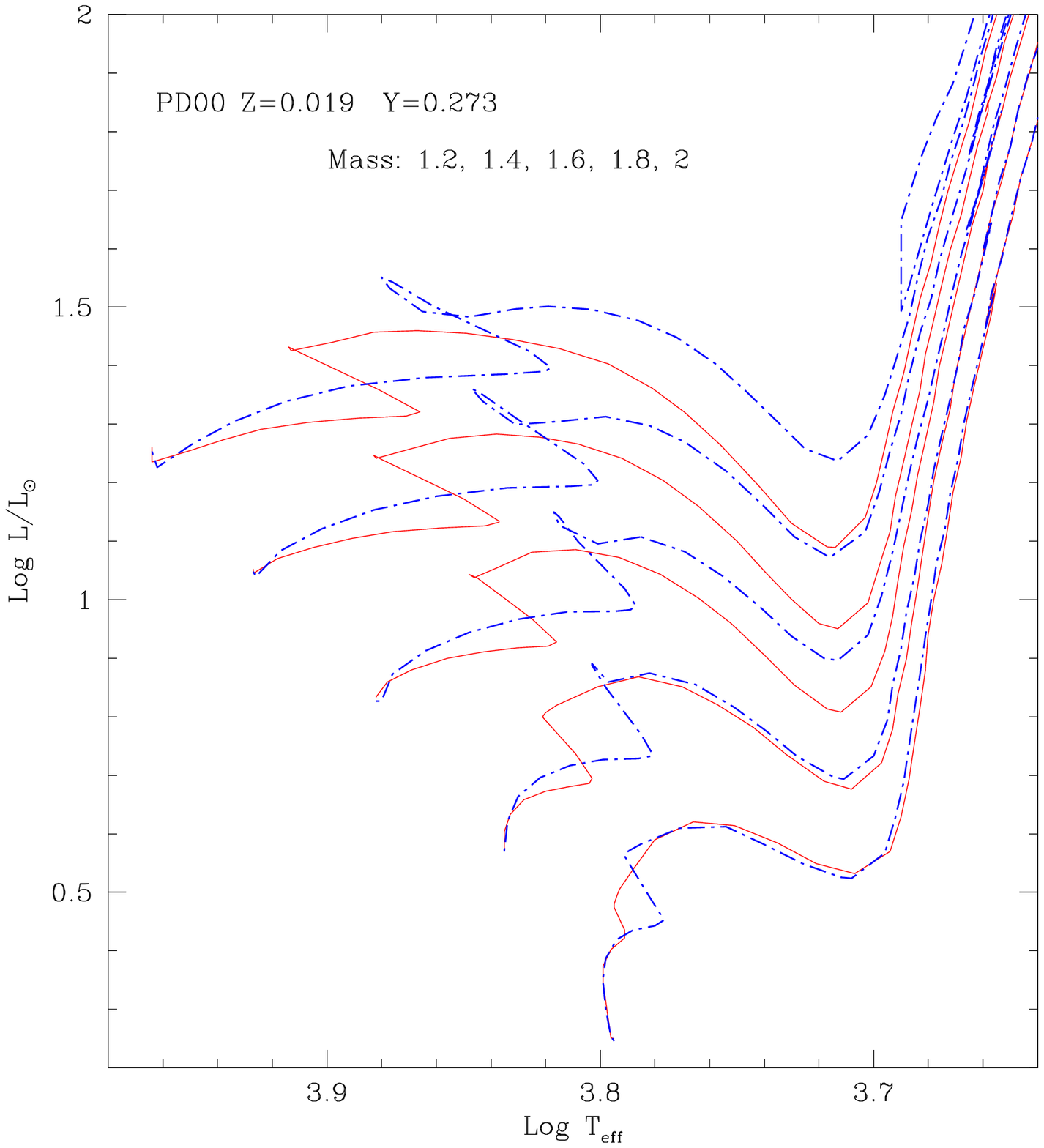}}
\end{minipage} 
\hfill
\begin{minipage}{0.45\textwidth} \noindent b)
\resizebox{\hsize}{!}{\includegraphics{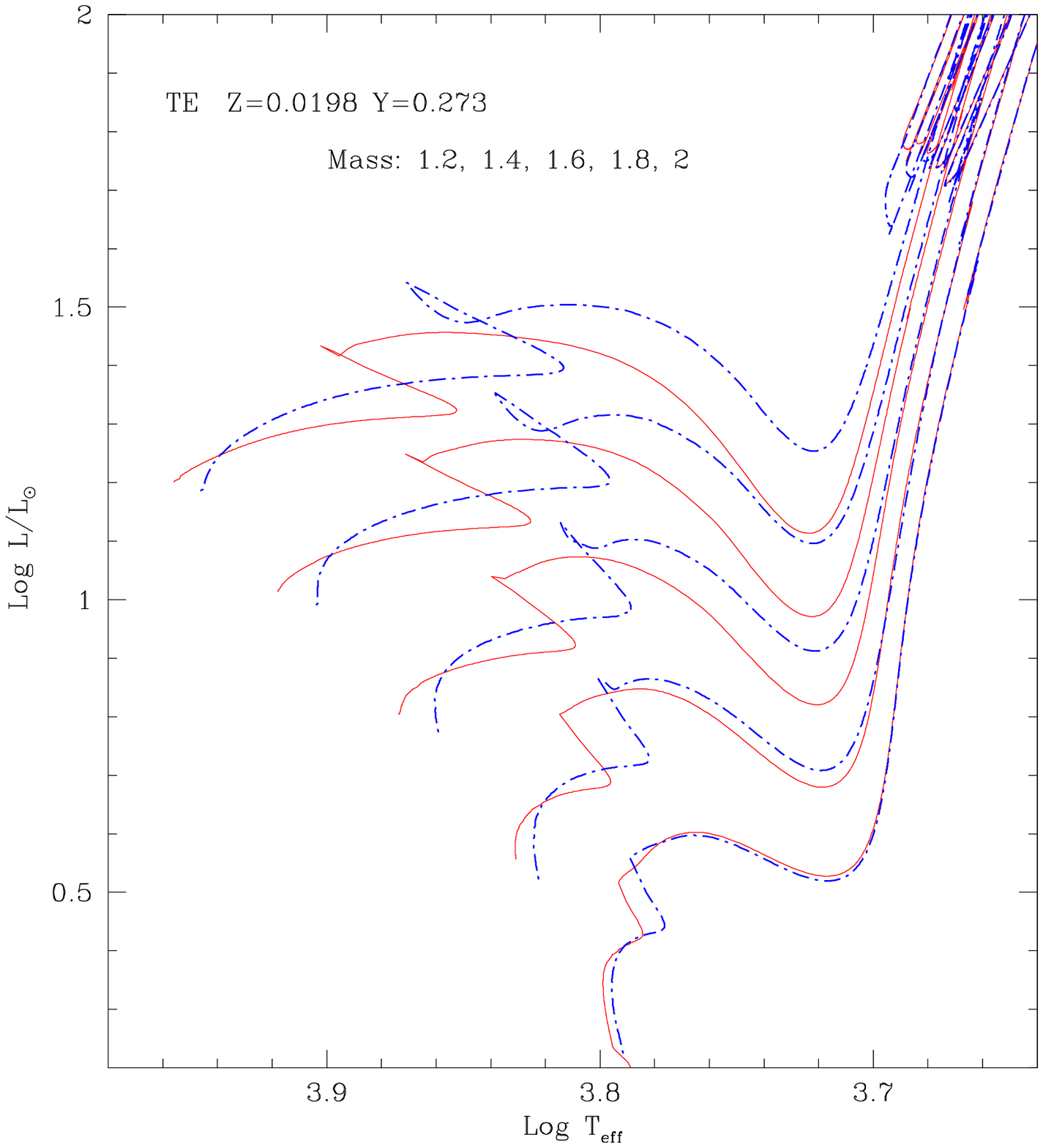}}
\end{minipage} 
\caption{  
Evolutionary tracks from Girardi et al. (2000)  for the composition $[Z=0.019,
Y=0.273]$ with overshoot (dot-dashed line) and without (solid line) in panel 
a).  In panel b) Teramo tracks are displayed for $[Z=0.0198, Y=0.273]$ with
overshoot (dot-dashed line)  and without (solid line).
 The ZAMS location for TE models with and 
without overshoot is not coincident.}
\label{PD_TEzsun}
\end{figure*} 

\begin{figure}
 \resizebox{\hsize}{!}{\includegraphics{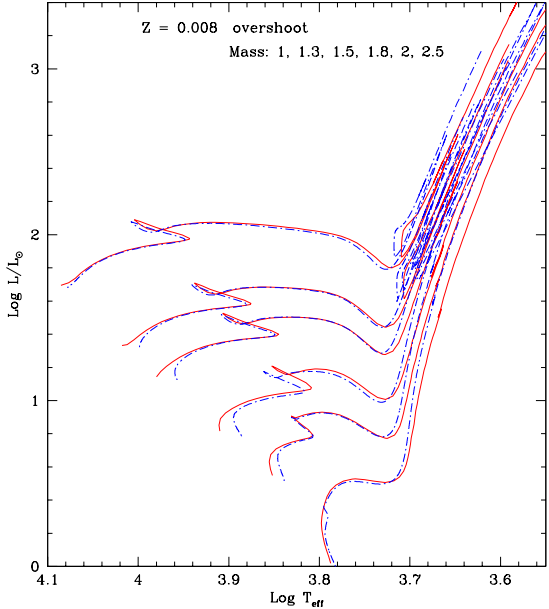}}
\caption{New Padova tracks and Teramo ones with overshoot from 1 to 2.5 
$M_{\odot}$ for Z=0.008.
Solid lines correspond to PD08 models, dash-dotted ones to Teramo tracks.}
\label{PD07_TEz008}
\end{figure}

\begin{table*}
\begin{center}
\caption{Comparison of H-lifetimes for models with overshoot with nearly solar composition and for Z=0.008 }
\label{tab_tH}
\begin{tabular}{lllllllll}
\noalign{\smallskip}\hline\noalign{\smallskip}
\multicolumn{4}{c}{$Z\sim Z_{\odot}$}& &\multicolumn{4}{c}{$Z=0.008$}\\
\noalign{\smallskip}
\cline{1-4}\cline{6-9}
\noalign{\smallskip}
$M/\Msun$  & $t_H$ TE $^1$  &  $t_H$ PD08 $^2$ & $t_H$ $YY$ $^3$ &  &
$M/\Msun$  & $t_H$ TE $^4$  &  $t_H$ PD08 $^5$ & $t_H$ $YY$ $^6$  \\
\noalign{\smallskip}\cline{1-4}\cline{6-9}\noalign{\smallskip}
  1.00 &  10.889  & 10.547 & 10.855 & & 1.00 & 8.387 & 7.389 & 7.795 \\
  1.20 &   5.679  &  5.109 &  5.359 & & 1.20 & 4.625 & 3.743 & 3.994 \\
  1.40 &   4.182  &  3.390 &  3.424 & & 1.30 & 4.009 & 3.040 & 3.076 \\
  1.60 &   3.067  &  2.347 &  2.390 & & 1.50 & 2.986 & 2.162 & 2.215 \\
  1.80 &   2.136  &  1.647 &  1.671 & & 1.80 & 1.847 & 1.266 & 1.306 \\
  2.00 &   1.439  &  1.198 &  1.229 & & 2.00 & 1.250 & 0.943 & 0.976 \\
  3.00 &   0.375  &  0.394 &  0.392 & & 3.00 & 0.331 & 0.324 & 0.334 \\
  5.00 &   0.094  &  0.107 &  0.104 & & 5.00 & 0.089 & 0.097 & 0.098 \\
\noalign{\smallskip}\hline\noalign{\smallskip}
\end{tabular}
    
$^1$  Z=0.0198 Y=0.273; $^2$  Z=0.017 Y=0.26; $^3$  Z=0.020 Y=0.27 

$^4$  Z=0.008  Y=0.256; $^5$  Z=0.008 Y=0.26; $^6$  Z=0.007 Y=0.244

Ages are in Gyr

\end{center}
\end{table*}

Figure 11 shows the comparison between the isochrones by Girardi
et al. (2000), hereafter PD00, with the new ones, hereafter PD08, in the
age range  between 1 and 14 Gyr, for the chemical composition $Z=0.017$ and
$Y=0.273$. There are
differences both in the MS phase and in the luminous part of the RGB
and/or AGB phase. The main contribution to the differences in the MS 
phase is due to the interpolation technique for the opacity tables (see Section
2.2). The amount of the variation depends on mass and chemical
composition. We have verified that low-mass stellar evolutionary tracks 
change very little with the method of opacity interpolation. For
intermediate mass stars there is a lowering of the luminosity 
and a consequent increase of the lifetime of the new models.
For values of log Age$ \le 9.5-9.7$ (it depends on
the chemical composition) to obtain the same luminosity of PD00
for the MS termination point in the new isochrones (PD08) we must increase 
the logarithm of the age  of about 0.05.  The color of the 
turn-off of the new isochrones (younger than 9.5 Gyr) is bluer.

We point out that the luminous part of the RGB and/or AGB phases in the new 
isochrones
is redder than in PD00 ones and the separation increases for increasing 
luminosity. This effect is due to the treatment of the 
density inversion in the red giant envelope. In the outermost layers of stars
with $T_{eff} \le  10^4 K$ a convective zone develops caused by the 
incomplete ionization of hydrogen. If convection is treated according to the 
mixing-length
theory (MLT) with the mixing-length proportional to the pressure scale height
($H_p$), when $T_{eff} \le  10^4 K$ a density inversion occurs. This situation 
might be unphysical and it has been the subject of many speculations. 
This difficulty can be avoided by adopting the density scale height $H_{\rho}$ 
instead of $H_p$ in the MLT. 
Really in Girardi et al. tracks we avoided the inversion density 
imposing that the temperature gradient  $\Delta T$ is such that the density 
gradient obeys the condition $\Delta _{\rho} \ge 0$. However the comparison of 
the PD00 isochrones with the observations seems to require redder RGB and/or 
AGB branches.
For this reason in the present computations we have used the MLT 
proportional to the pressure scale height, 
allowing the inversion density in the red convective envelope, as already
adopted in Bertelli et al. (1994) isochrones.  The
presence of the inversion density determines RGB and/or AGB phases redder 
than in previous isochrones (PD00).  

Neutrino cooling is also more efficient during the RGB phase in the new tracks,
as described in Section 2.4 and it contributes to make the tip of the RGB more 
luminous than in PD00. We have verified that for Z in the range 0.0001 - 0.001 
(low values of Y) and for the solar case the overluminosity is of
the order 0.12 - 0.15 in  units of $log L/L_{\odot}$ (the considered ages are 
around  log Age=10.0).
The effect on the I magnitude consists in an overluminosity of the order
of between -0.20 and -0.35 for Z in the range 0.0001 - 0.001. On the other hand
 in  the solar case
the interplay between lowering effective temperature and increasing
bolometric corrections is such that the tip of the PD00 case appears to be
approximately one magnitude more luminous than the new one, inverting 
what happens to the I magnitude for low values of Z. 

The comparison of some selected tracks and isochrones of the new grids  
 of  the Padova database(PD08) with the analogous
counterparts of the publicly available Teramo database (TE) by Pietrinferni et 
al. (2004) and the $YY$ database by Yi et al. (2001) and Demarque et al. 
(2004) pointed out some differences.

As already evident in the comparison of isochrones 
by different groups, shown in Pietrinferni et al. (2004), there is a systematic
difference from their 1.8 Gyr isochrone and the one by Girardi et al. (2000) 
and by Yi et
al. (2001), as for the 2.0 Gyr one by VandenBerg (see figures 8, 11 and 12 in
Pietrinferni et al. 2004). In fact the turn-off 
luminosity of their 1.8 Gyr isochrone (or 2.0 in the comparison with 
VandenBerg) is higher, whereas the 10 Gyr and the 0.5 Gyr
ones are practically the same as by the other groups. They ascribed the 
differences to differences in input physics and in the overshooting 
treatment relatively to the comparison with Girardi et al. (2000).

In Table 4 we compare the H-lifetimes for models with overshoot in 
the mass range between 1 and 5 $M_{\odot}$ and with a chemical composition
approximately solar 
We point out that the largest differences can be found in the range
between 1 and 2 solar masses, in the sense that while PD08 and $YY$ tracks 
have similar H-lifetimes, TE models are systematically longer up to 0.8 Gyr.
This difference cannot be ascribed to the treatment of overshoot, in fact
both groups (PD and TE) assume a similar prescription in the range between
1 and 2 $M_{\odot}$ (the core overshoot efficiency decreases linearly from 
about $1.7-1.5 M_{\odot}$ down to $1 M_{\odot}$, where the efficiency is
zero for stars with a radiative core). 

The recent determination of the age of the LMC globular cluster NGC 1978,
studied by Mucciarelli et al. (2007) shows the same anomaly, in this case 
for the chemical composition typical of the LMC. The authors find that the
age obtained with Teramo isochrones results significantly older than that 
obtained with Pisa and Padova isochrones, even if the same amount of overshoot
is assumed in the considered models. 
Comparing the H-lifetimes for $Z=0.008$ we notice that there is a very 
satisfactory agreement between the H lifetimes of $YY$ and PD08, whereas TE 
ages are older up to about 1 Gyr at about $1.3 M_{\odot}$. The largest 
differences are present in the range between 1 and 2 $M_{\odot}$, as shown in
Table~\ref{tab_tH}.

Figure 12 panel a) shows isochrones (at log Age = 10.0, 9.7, 
9.4, 9.2, 9.0 years) from
Teramo database with Z=0.008 and Y=0.256, and (interpolated at the same metal
and helium content) from our new isochrones. This figure remarks the difference
in luminosity of the isochrones for the same age from the two databases, 
more evident for log Age=9.4 (black solid line for PD08 and black dash-dotted 
for Teramo) in the color version of the electronic paper. 
In panel b) we compare Padova new isochrones for Z=0.007 and 
Y=0.244  with $YY$  and we
find out that the agreement is very good, considering that their
models take into account convective core overshoot (0.2) and helium diffusion
(PD08 and $YY$ isochrones are shown for log Age = 10.0, 9.7, 9.4 and 9.0 
years).

The reasons of the above mentioned disagreement cannot be easily disentangled, 
but we notice that the ZAMS models for PD00 tracks (Girardi et al. 2000 for
solar chemical composition) with and without overshoot are coincident (as the
ZAMS is the location of chemically homogeneous models for the various 
considered masses), while they are different in luminosity and temperature
for Teramo ones between 1 and 2 $M_{\odot}$.  
 The difference in the location of ZAMS models can be seen also
in Figure 14 where we plot our new models and Teramo ones
with overshoot for Z=0.008.

In Figure 13 the PD00 and TE tracks are plotted in the range of 
mass between 1.2 and 2 $M_{\odot}$. As we did not compute new tracks without
overshoot, we plotted PD00 and TE models with and without overshoot to make
a comparison. 
The anomalously longer H-burning lifetimes of TE models with respect 
to other authors is probably, at least partly, related to the lower luminosiy 
of the beginning of the central H-burning for their models with
overshoot.
We point out that Teramo models with overshoot were computed
starting from the pre-MS phase for masses lower than $3 M_{\odot}$.
 Anyway also $YY$ models were evolved from the pre-MS stellar birthline to
the onset of helium burning (Yi et al. 2001), but there are not significant
differences (comparing PD08 to YY results) in the  H-lifetimes and for the 
isochrones in the age range between 1 and 10 Gyr, as shown  in Table 4 and 
Figure 12 b).

%%%%%%%%%%%%%%%%%%%%%%%%%%%%%%%%%%%%%%%%%%%%%%%%%%%%%%%%%%%%%%%%%%%%%
\section{Concluding remarks}
\label{sec_remarks}

%%%%%%%%%%%%%%%%%%%%%%%%%%%%%%%%%%%%%%%%%%%%%%%%%%%%%%%%%%%%%%%%%%%%%

Large grids of homogeneous stellar evolution models and isochrones are a 
necessary ingredient for the interpretation of photometric and spectroscopic 
observations of resolved and unresolved stellar populations.
Appropriate tools to compute synthetic color-magnitude diagrams for the 
analysis of stellar populations require also the selection of the chemical
composition to be used in the simulation. Evolutionary stellar models are
usually computed taking into account a fixed law of helium to metal enrichment.
Determinations of the helium enrichment $\Delta Y / \Delta Z$ from nearby stars
and K dwarfs, or
from Hyades binary systems show a large range of values for this ratio.
Recent results suggest that the naive assumption that the helium enrichment
law is universal, might not be correct.
In fact there has been recently evidence of significant variations in the 
helium content (and perhaps of the age) in some globular clusters, like 
$\omega$ Cen, NGC 2808 and NGC 6441, while 
globular clusters were traditionally considered as formed of a simple
stellar population of uniform age and chemical composition.  

These results prompted us to compute new stellar evolutionary tracks covering
an extended region of the plane Z-Y, in order to enable users to analyse 
stellar populations with different helium enrichment laws.

 In this paper we present 39 grids of stellar evolutionary tracks up to 
$2.5 M_{\odot}$. Next paper will present tracks and isochrones from $2.5$
up to $20 M_{\odot}$ for the same grid of chemical
compositions. The typical mass resolution for low mass stars is of 0.1 
$M_{\odot}$, reduced to $0.05$ in the interval of very low masses ($M < 0.6
M_{\odot}$), and occasionally in the vicinity of the separation mass $M_{Hef}$
between low and intermediate mass stars. In this way we provide enough tracks 
to allow a very detailed mapping of the HR diagram and the relative theoretical
isochrones are also very detailed. An important update of this database is the
extension of stellar models and isochrones until the end of the TP-AGB
 by means of new synthetic models (cf. Marigo \& Girardi 2007).
The results of these new grids allow the estimate of several astrophysical 
quantities as the RGB transition mass, the He core mass at He ignition, the 
amount of changes of chemical elements due to the first dredge-up, and the 
properties of the models at the beginning of the thermal pulse phase on the 
AGB.

The web site (http://stev.oapd.inaf.it/YZVAR), dedicated to make available the
whole theoretical framework to the scientific community, includes:
\begin{itemize}
\item
{Data files with the information relative to each evolutionary track}
\item
{Isochrone files with the chemical composition of the computed grids in 
the Z-Y plane  }
\end{itemize}
Every file contains isochrones with $\sim 10.15 \le$ Log Age $ \le \sim 9.0$ 
years with
time step equal to 0.05. The initial and the final value can be a little more 
or a little less, dependent on the chemical composition.
To compute the isochrones we adopted the interpolation scheme described in 
Section 5.1. A web interactive interface will be tuned up as soon as possible
to allow users to obtain isochrones of whatever chemical composition inside
the original grid.

A modified version of the program which computes the isochrones is suited to
the purposes of evolutionary population synthesis, either of simple (star
clusters) or complex stellar populations (galaxies). In the latter case it is 
possible to select the kind of star formation history, the chemical enrichment
law in the plane Z-Y, the initial mass function and the mass loss rate during 
the RGB phase. The previous program ZVAR, in which the helium enrichment law
was implicit in the chemical composition of the various evolutionary sets, is 
now updated for being used in the extended region of the Z-Y plane and is named
YZVAR.
A web interactive interface will provide stellar populations for selected input
data for the SFR, IMF, chemical composition and mass loss during the RGB phase.

\begin{acknowledgements}
We thank  C. Chiosi for his continuous interest and support to stellar
evolution computations. We thank A. Weiss and B. Salasnich for help with 
opacity tables, and A. Bressan for useful discussions.
The authors acknowledge constructive comments of the referee that helped
to clarify and improve the text.  
We acknowledge financial support from INAF COFIN 2005 ``A Theoretical lab
for stellar population studies'' and  from
Padova University (Progetto di Ricerca di Ateneo CPDA 052212).

\end{acknowledgements}

%%%%%%%%%%%%%%%%%%%%%%%%%%%%%%%%%%%%%%%%%%%%%%%%%%%%%%%%%%%%%%%%%%%%%

\end{document}